\documentclass[reprint,amsmath,amssymb,aps,prb]{revtex4-2}
\usepackage[colorlinks=true,citecolor=blue,linkcolor=red,filecolor=blue,urlcolor=blue]{hyperref} 

\usepackage{graphicx}
\usepackage{dcolumn}
\usepackage{bm}
\usepackage{color}
\usepackage{ulem}

\begin{document}

\title{
Entropic crystallization of geometrically frustrated magnets on 1/1 approximant Tsai-type quasicrystal
}

\author{Oscar Novat$^{1,2}$, Ludovic D. C. Jaubert$^{3}$, and Masafumi Udagawa$^{2}$}%
\affiliation{%
$^1$ENS de Lyon, CNRS, Laboratoire de Physique, F-69342 Lyon, France\\
$^2$Department of Physics, Gakushuin University, Mejiro, Toshima-ku, Tokyo 171-8588, Japan\\
$^3$CNRS, Universit\'{e} de Bordeaux, LOMA, UMR 5798, 33400 Talence, France
}%

\date{\today}

\begin{abstract}
We have studied the antiferromagnetic Ising model on the icosahedral bcc lattice, as a model system of 1/1 approximant Tsai-type quasicrystals.
We addressed thermal equilibrium properties of this system with Markov-chain Monte Carlo simulation supplemented with the parallel tempering technique to accelerate the relaxation dynamics.
As a result, we found a second-order phase transition takes place to the magnetic ordered phase with ${\mathbb Z_3}\times {\mathbb Z_2}$ symmetry breaking.
Despite the ordering, the low-temperature phase keeps macroscopic degeneracy as identified by finite residual entropy, $\mathcal{S}\sim0.1767/{\rm spin}$.
Remarkably, the existence of residual entropy turns out to play a major role in the formation of magnetic order. 
Generation of domain wall is suppressed, as it reduces the residual entropy locally stored in icosahedra, beyond the gain of configurational entropy due to domain wall patterns.
Magnetic order arises out of this competition as entropic crystallization, which manifest universal mechanism of strongly frustrated systems with large geometrical units.
\end{abstract}

\maketitle

\section{Introduction}
Geometrical frustration has produced a rich variety of exotic magnetic states~\cite{UdagawaJaubert2021,ANDERSON1973153,PhysRevLett.86.1881,kitaev2006anyons,ANDERSON1978291,PhysRevLett.70.3339}. 
Among the frustrated lattice structures, considerable attention is focused on the lattices made of corner-sharing units -- such as kagome, checkerboard, trillium and pyrochlore -- whose physics is more easily understood in two steps; first within a triangular or tetrahedron unit, and then in the lattice as a whole. Among these systems, spin ice is probably the most famous example~\cite{UdagawaJaubert2021}.
Defined on a corner-sharing network of tetrahedra, it exhibits the unconventional properties of a classical Coulomb spin liquid with macroscopic residual entropy~\cite{Ramirez:1999aa}, fractionalized magnetic monopoles~\cite{castelnovo2008magnetic} and magnetic correlation with pinch point singularities~\cite{doi:10.1126/science.1177582}. Intensive studies are going on to further clarify various aspects of spin ice, such as classical and quantum dynamics of magnetic monopoles~\cite{PhysRevLett.122.117201,tokiwa2016tokiwa,Poree:2025aa,PhysRevLett.124.097204,PhysRevB.96.085136,doi:10.7566/JPSJ.90.123705}, magnetization plateaus~\cite{doi:10.1143/JPSJ.71.2365,Matsuhira:2002aa,PhysRevB.68.064411,Tang:2023aa,kermarrec2025magnetizationmagnetostrictionmeasurementsdipoleoctupole}, diffusive magnetic correlations~\cite{PhysRevB.94.104416,PhysRevLett.119.077207,rau2016spin,mizoguchi2018magnetic,PhysRevB.109.174421,PhysRevB.110.195117}, magneto-elastic coupling~\cite{Khomskii:2012aa,Khomskii:2021aa,Uehara:2022aa}, glassy slow dynamics in absence of any structural disorder~\cite{Matsuhira:2000aa,Matsuhira:2001aa,Snyder:2001aa,Jaubert:2009aa}, and thermal transport~\cite{PhysRevLett.110.217209,PhysRevB.86.060402,PhysRevB.88.054406,PhysRevB.105.104405,tang2025observationspinseebeckeffect}.

\begin{figure}[ht]
\begin{center}
\includegraphics[width=0.35\textwidth]{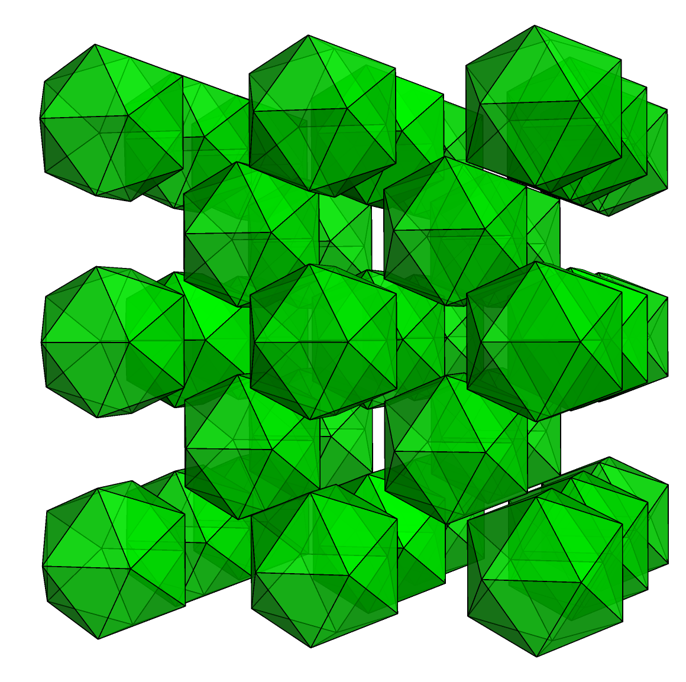}
\end{center}
\caption{\label{Fig1} 
(Color online) The icosahedral bcc network of magnetic ions on the 1/1 quasicrystal approximant.}
\end{figure}

Here we consider the approximants of quasicrystals as extensions of these paradigmatic frustrated lattices. Approximants are periodic lattices whose structure locally approximates the atomic arrangement of a quasicrystal.
Among them, the so-called 1/1 approximant takes a relatively simple lattice structure, where a basic geometrical unit has rhombic triacontahedral structure, which consists of successive shells made of an icosidodecahedron, an icosahedron, a dodecahedron, and a tetrahedron, from outer to inner shell~\cite{Takakura:2007aa,PhysRevB.68.024203}. 
In a typical class of Yb-Cd 1/1 approximant Tsai-type quasicrystals, magnetic Yb ions are placed on the icosahedral shell of the unit. In other words, if only magnetic degrees of freedom are considered, the lattice structure of a 1/1 approximant can be simplified as a set of icosahedra placed on the vertices of the body-centered-cubic (bcc) lattice [Fig.~\ref{Fig1}]. 
Hence, this class of magnetic approximant quasicrystal can be regarded as a generalization of spin ice, where icosahedra, instead of tetrahedra, are the source of geometrical frustration.

On the experimental front, many 1/1 approximant crystals have been synthesized and their magnetic properties are studied~\cite{PhysRevB.98.220403,PhysRevB.82.220201,doi:10.1143/JPSJ.81.024720,doi:10.7566/JPSJ.86.093702,Ibuka_2011,PhysRevB.93.024416,PhysRevB.100.180409,Hiroto_2013,Hiroto_2014,PhysRevB.88.214202,PhysRevB.101.180405}.
In the classification table of Ref.~\cite{ShintaroSuzuki2021MT-MB2020014}, a substantial amount of magnets are dubbed as ``spin glass".
While there is a possibility of intrinsic spin-glass physics in these materials, this glassiness suggests that the magnetic properties of several compounds are still unclear.
This uncertainty may be attributed to a shortage of theoretical studies on quasicrystal approximants, leaving many potential magnetic patterns unexplored.

To help resolve the magnetic structures of realistic materials, in this work, we aim to find typical magnetic patterns of 1/1 approximant Tsai-type quasicrystals, induced by geometrical frustration, by considering the antiferromagnetic Ising model.
For the systems with ferromagnetic interaction, many interesting magnetic patterns have been proposed, and their thermodynamic and transport properties have been studied~\cite{Watanabe:2021aa,PhysRevB.109.184404,vqvs-rbzn,4zb8-2zjq,doi:10.7566/JPSJ.90.063701,doi:10.7566/JPSJ.95.044705}.
Compared with the intensive efforts on ferromagnetic side, the number of studies on frustrated interactions is still limited~\cite{Jeon:2024aa,doi:10.7566/JPSJ.85.053701,doi:10.7566/JPSJ.93.045001}.
In this work, frustration gives rise to the co-existence of long-range magnetic order specific to the structure of icosahedral units on one hand, together with fluctuating spins supporting a finite residual entropy down to zero temperature, similarly to classical spin liquids.
Surprisingly, this residual entropy turns out to be a driving force of magnetic order through the competition between local icosahedral entropy and global configurational entropy of domain wall patterns.

\section{Model \& Method}

\subsection{Model}
\label{sec:model}

\begin{figure}[ht]
\begin{center}
\includegraphics[width=0.5\textwidth]{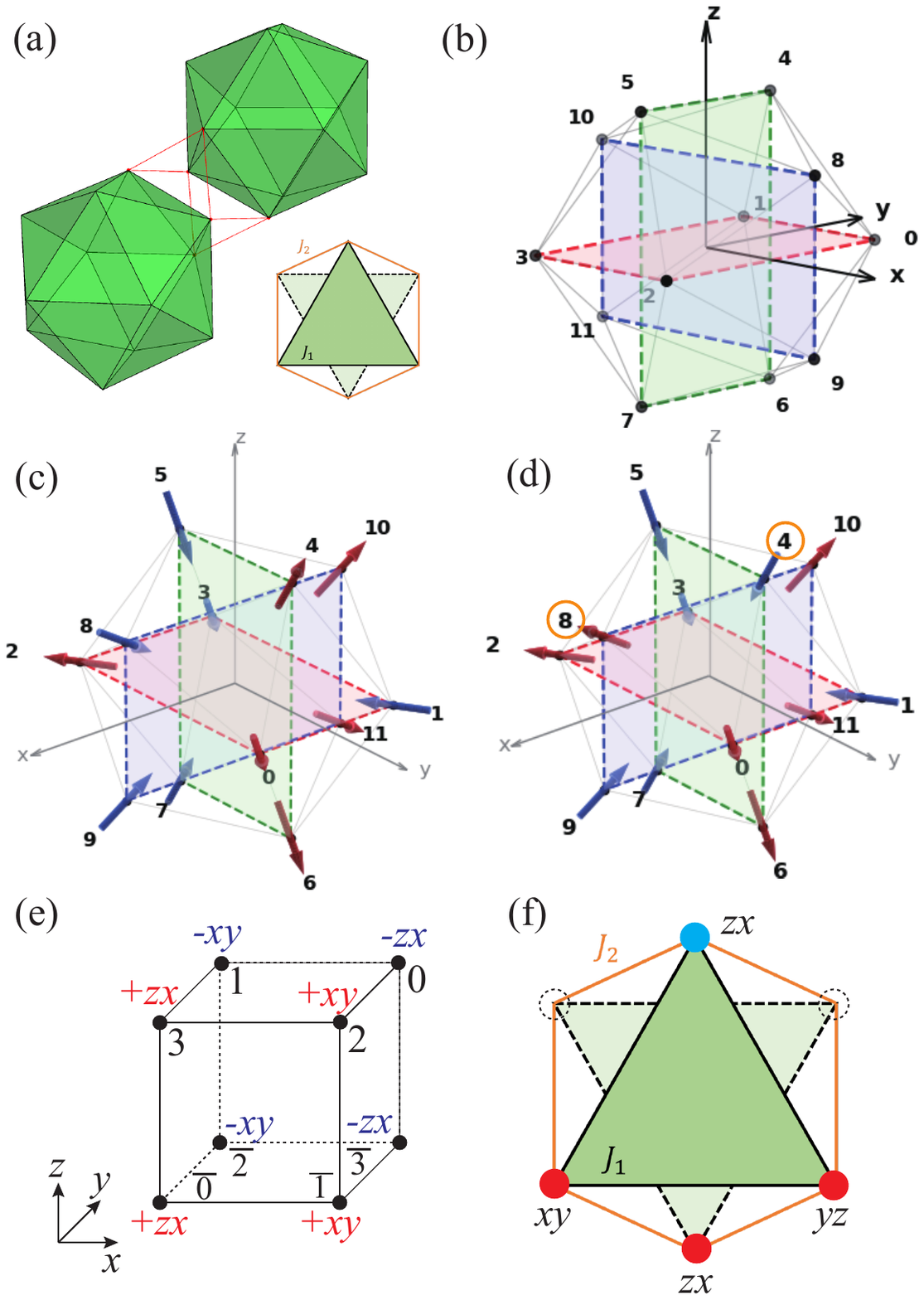}
\end{center}
\caption{\label{Fig2} 
(Color online) (a) Structure of the connecting triangle. (b) Numbering of sites within an icosahedron. Rectangles on each plane are shown. (c) The spin configuration of the P1 state of the XY sector is shown (see Table \ref{tab:table3}). $\sigma=+1(-1)$ is shown with outward (inward) arrow. (d)  The spin configuration of the S1 state of the XY sector is shown (see Table \ref{tab:table3}). The spin configuration is obtained by exchanging spins at sites $4$ and $8$ from the P1 state as shown with orange circles. (e) The positions of the 8 connecting triangles within an icosahedron are shown on the vertices of a cube. The spin minority representation of the P1 state is shown.
(f) Two connecting triangles belonging to two neighboring icosahedra. The top triangle has a minority spin $\sigma=-1$ on the $zx$ plane, as colored in blue. Due to the ferromagnetic $J_2$, the $zx$ spin of the bottom triangle takes $\sigma=+1$, as colored in red.}
\end{figure}

We consider the $J_1$-$J_2$ Ising model defined on the icosahedral bcc lattice, where an icosahedron is placed at each vertex of the bcc lattice [Fig.~\ref{Fig1}].
The Hamiltonian reads
\begin{eqnarray}
\mathcal{H} = \mathcal{H}_1 + \mathcal{H}_2 = J_1\sum_{\langle i,j\rangle}\sigma_i\sigma_j + J_2\sum_{\langle\langle i,j\rangle\rangle}\sigma_i\sigma_j.
\label{eq:Hamiltonian}
\end{eqnarray}
Here, the Ising variable takes $\sigma_j=\pm1$. We assume the antiferromagnetic coupling inside each icosahedron, $J_1\equiv 1$, which we take as a unit of energy.
We also introduce the coupling $J_2$ between neighboring icosahedra.
On the bcc lattice, neighboring icosahedra face with each other through a pair of triangles [Fig.~\ref{Fig2} (a)].
We define the $J_2$ coupling between the nearest-neighbor sites on this connecting triangle pair. Note that, as opposed to kagome and pyrochlore lattices, our system is not made of corner-sharing frustrated units. Each Ising spin belongs to one and only one icosahedron, and interacts with four spins of two neighboring icosahedra via $J_2$ coupling.

We arbitrarily choose $J_2$ to be ferromagnetic, $J_2\leq0$, because the sign of $J_2$ is irrelevant to the physics. Since the bcc lattice is bipartite, 
if we replace $\sigma_j$ with $-\sigma_j$ on all icosahedra for one of the two bcc sublattices, then the system can be mapped to the model with antiferromagnetic $J_2$.

\subsection{Monte Carlo simulation}
We address the finite temperature properties of Hamiltonian (\ref{eq:Hamiltonian}) with classical Monte Carlo simulation.
We use the standard single spin flip Metropolis algorithm, jointly with parallel tempering in order to facilitate thermalization. However, as is often the case with geometrically frustrated systems, simulations encounter dynamical bottlenecks of relaxation.
In particular, each vertex of an icosahedron has an odd number of nearest-neighbor sites.
This means that the effective field is always non-zero, so the single spin flip process eventually becomes frozen at low temperatures.
To remedy this problem, we also introduce a two-spin update which can take place at zero energy cost.

The simulation runs on $N$ parallel cores; each core stores one temperature for a cubic system of $N_{\rm spin}=12\times2\times L^3$ spins.
The $12$ is for the twelve spins in one icosahedron, the $2$ is for the two sublattices of the bcc lattice, and $L$ counts the number of bcc unit cells along one of the cartesian axes.
One Monte Carlo step is the collection of $N_{\rm spin}$ single and double spin flips, chosen randomly on the lattice and accepted through a detailed balance condition.
Each core undergoes an annealing procedure before the simulation starting at $T_i+500$, where $T_i$ is the initial temperature of the simulation (the greatest among cores). 
This annealing procedure is the same for all cores: the interval $\left[T_i+500,T_{\rm core}\right]$ is divided into a linear slope of $10^6$ temperature steps, completing one Monte Carlo step at each temperature.

Each core is then thermalized during $10^6$ Monte Carlo steps at its own temperature $T_{\rm core}$. 
We do not perform any measurements during this thermalisation phase, but we use parallel tempering between the cores every $10$ Monte Carlo steps.

Finally, the system is sampled into $N_{\rm sample}$ different samples to perform the thermal averages.
Each sample is obtained after $10$ Monte Carlo steps, and parallel tempering is applied  every $100$ Monte Carlo steps.
In all the results shown in this article, $N_{\rm sample}$ is set to $10^6$.
The parameters that can be varied along simulations are $J_2$ and $L$.

\section{The ground state of a single icosahedron}

\subsection{Degeneracy}
At $J_2=0$, the system is decomposed into independent icosahedra, where each icosahedron has 72-fold degenerate ground states.
To confirm this, it is easiest to rewrite the Hamiltonian as the summation over triangles.
\begin{eqnarray}
\mathcal{H} = J_1\sum_{\langle i,j\rangle}\sigma_i\sigma_j = \frac{J_1}{4}\sum_{i,j,k\in\triangle}[(\sigma_i+\sigma_j+\sigma_k)^2 -3].
\label{eq:SquaredHamiltonian}
\end{eqnarray}
According to this expression, we recover the traditional constraint on antiferromagnetic triangular structures that imposes $\sigma_i+\sigma_j+\sigma_k=\pm1$  on all $N_{\triangle}\equiv$ 20 triangles of the icosahedron. For the sake of convenience, we shall refer to this constraint as the triangle condition. Consequently, the ground state energy is $E=-\frac{J_1}{2}N_{\triangle}=-10J_1$ per icosahedron.
By simple arguments based on local effective fields, one finds that the triangle condition indeed leads to 72 ground states per icosahedron, as derived in Appendix \ref{Appendix:GSsingleicosahedron}.

\subsection{Sectors}
\label{sec:Sector}
The 72 configurations of the icosahedral ground state can be classified into 6 sectors.
To see this, we first define the spin positions within an icosahedron, as shown in Fig.~\ref{Fig2} (b).
An icosahedron can be decomposed into three rectangles, each parallel to $xy$, $yz$, and $zx$ planes.
As explicitly shown in Fig.~\ref{Fig2} (b), spins are numbered so that $(\sigma_0, \sigma_1, \sigma_2, \sigma_3)$, $(\sigma_4, \sigma_5, \sigma_6, \sigma_7)$, and $(\sigma_8, \sigma_9, \sigma_{10}, \sigma_{11})$ are on the $xy$, $yz$, and $zx$ planes, respectively. For the $\{x,y,z\}$ coordinates of each spin, see Table~\ref{tab:table1}.

\begin{table}[t]
    \centering
    \begin{minipage}{0.45\linewidth}
    \begin{tabular}{|c|c|}
        \hline
        Site & $xyz$-coordinate \\
        \hline
        0 & $(+1, +\tau, 0)$\\
        1 & $(-1, +\tau, 0)$\\
        2 & $(+1, -\tau, 0)$\\
        3 & $(-1, -\tau, 0)$\\      
        \hline
        4 & $(0, +1, +\tau)$\\
        5 & $(0, -1, +\tau)$\\
        6 & $(0, +1, -\tau)$\\
        7 & $(0, -1, -\tau)$\\  
        \hline
        8 & $(+\tau, 0, +1)$\\
        9 & $(+\tau, 0, -1)$\\
        10 & $(-\tau, 0, +1)$\\
        11 & $(-\tau, 0, -1)$\\ 
        \hline
    \end{tabular}
    \end{minipage}
     \begin{minipage}{0.45\linewidth}
    \begin{tabular}{|c|c|}
        \hline
        Triangle & Sites ($xy$, $yz$, $zx$) \\
        \hline
        0 & $(0, 4, 8)$\\
        1 & $(1, 4, 10)$\\
        2 & $(2, 5, 8)$\\
        3 & $(3, 5, 10)$\\     
        \hline         
        $\overline{0}$ & $(3, 7, 11)$\\ 
        $\overline{1}$ & $(2, 7, 9)$\\
        $\overline{2}$ & $(1, 6, 11)$\\
        $\overline{3}$ & $(0, 6, 9)$\\             
        \hline
    \end{tabular}
    \end{minipage}
    \caption{\footnotesize (Left) $xyz$-coordinates of the sites on a single icosahedron. $\tau\equiv\frac{1+\sqrt{5}}{2} > 1$ is a golden ratio. Accordingly, the bond $02$ is longer than $01$. The upper, middle and bottom 4 rows are on the $xy$, $yz$, and $zx$ planes, respectively. (Right) The triplet of spins $(\sigma_i, \sigma_j, \sigma_k)$ constituting each connecting triangle. $\sigma_i$, $\sigma_j$, and $\sigma_k$ are on the $xy$, $yz$, and $zx$ planes in this order. The connecting triangle $\bar{a}$ is at the opposite side of $\bar{a}$ ($a=0,1,2,3$). See also Fig.~\ref{Fig2}(d)}
    \label{tab:table1}
\end{table}

Among the 72 ground states, 24 possess the property of carrying zero magnetization on all three rectangles. These 24 states can be divided in a unique way into 6 sectors of 4 states each. The condition is that, within any given sector, the 4 states share the same spin configuration on 2 of the 3 rectangles; it is only the third rectangle whose spin configuration distinguishes between the four states (see states P1 \dots P4 in Table \ref{tab:table3} for a given sector). Among the remaining $72-24=48$ ground states, there is always one, and only one, rectangle with zero magnetization. This rectangle shares the same configuration as in one of the six sectors mentioned above, which allows to divide these 48 remaining ground states into the 6 sectors (see states S1 \dots S4 in Table \ref{tab:table3}). The resulting sectors are thus made of 12 ground states each: (i) 4 of them with zero magnetization on all rectangles (we shall call them primary), and (ii) 8 of them with only one zero-magnetization rectangle (we shall call them secondary). Within a sector, each secondary state becomes a primary state by flipping two spins only [Fig.~\ref{Fig2}(c,d)].

The rectangle with zero magnetization for all states of a sector defines the principal plane of that sector. It enables to label the six sectors as XY, YZ, ZX and their time-reversal counterparts $\overline{\rm XY}$, $\overline{\rm YZ}$ and $\overline{\rm ZX}$. The spin configuration of that zero-magnetization rectangle always has opposite spins on the short side of the rectangle; the corresponding plane will be called the short-side-staggered (SSS) plane of the sector (see the red spins in Table \ref{tab:table3}). In parallel, the other rectangle with identical spin configuration for the primary state (see the green spins in Table \ref{tab:table3}) always have opposite spins on the long side of the rectangle; the corresponding plane will be called the long-side-staggered (LSS) plane of the sector.

\begin{table*}[ht]
    \centering
    \begin{tabular}{|c||c|c|c|c||c|c|c|c||c|c|c|c||c|}
        \hline
        Label & $\sigma_0$&$\sigma_1$&$\sigma_2$&$\sigma_3$&$\sigma_4$&$\sigma_5$&$\sigma_6$&$\sigma_7$&$\sigma_8$&$\sigma_9$&$\sigma_{10}$&$\sigma_{11}$ & Connections \\
        \hline
        P1 & \color{red}{$+$} & \color{red}{$-$} & \color{red}{$+$} & \color{red}{$-$} & $+$ & $-$ & $+$ & $-$ & \color{green}{$-$} & \color{green}{$-$} & \color{green}{$+$} & \color{green}{$+$} & $(4-8)(5-10)(6-9)(7-11)$\\
        P2 & \color{red}{$+$} & \color{red}{$-$} & \color{red}{$+$} & \color{red}{$-$} & $+$ & $-$ & $-$ & $+$ & \color{green}{$-$} & \color{green}{$-$} & \color{green}{$+$} & \color{green}{$+$} & $(4-8)(5-10)(6-11)(7-9)$\\
        P3 & \color{red}{$+$} & \color{red}{$-$} & \color{red}{$+$} & \color{red}{$-$} & $-$ & $+$ & $+$ & $-$ & \color{green}{$-$} & \color{green}{$-$} & \color{green}{$+$} & \color{green}{$+$} & $(4-10)(5-8)(6-9)(7-11)$\\
        P4 & \color{red}{$+$} & \color{red}{$-$} & \color{red}{$+$} & \color{red}{$-$} & $-$ & $+$ & $-$ & $+$ & \color{green}{$-$} & \color{green}{$-$} & \color{green}{$+$} & \color{green}{$+$} & $(4-10)(5-8)(6-11)(7-9)$\\
        \hline
        S1 & \color{red}{$+$} & \color{red}{$-$} & \color{red}{$+$} & \color{red}{$-$} & $-$ & $-$ & $+$ & $-$ & $+$ & $-$ & $+$ & $+$ & P1, P3 via spin 8\\
        S2 & \color{red}{$+$} & \color{red}{$-$} & \color{red}{$+$} & \color{red}{$-$} & $-$ & $-$ & $-$ & $+$ & $+$ & $-$ & $+$ & $+$ & P2, P4 via spin 8\\
        S3 & \color{red}{$+$} & \color{red}{$-$} & \color{red}{$+$} & \color{red}{$-$} & $+$ & $-$ & $-$ & $-$ & $-$ & $+$ & $+$ & $+$ & P1, P2 via spin 9\\
        S4 & \color{red}{$+$} & \color{red}{$-$} & \color{red}{$+$} & \color{red}{$-$} & $-$ & $+$ & $-$ & $-$ & $-$ & $+$ & $+$ & $+$ & P3, P4 via spin 9\\
        S5 & \color{red}{$+$} & \color{red}{$-$} & \color{red}{$+$} & \color{red}{$-$} & $+$ & $+$ & $+$ & $-$ & $-$ & $-$ & $-$ & $+$ & P1, P3 via spin 10\\
        S6 & \color{red}{$+$} & \color{red}{$-$} & \color{red}{$+$} & \color{red}{$-$} & $+$ & $+$ & $-$ & $+$ & $-$ & $-$ & $-$ & $+$ & P2, P4 via spin 10\\
        S7 & \color{red}{$+$} & \color{red}{$-$} & \color{red}{$+$} & \color{red}{$-$} & $+$ & $-$ & $+$ & $+$ & $-$ & $-$ & $+$ & $-$ & P1, P2 via spin 11\\
        S8 & \color{red}{$+$} & \color{red}{$-$} & \color{red}{$+$} & \color{red}{$-$} & $-$ & $+$ & $+$ & $+$ & $-$ & $-$ & $+$ & $-$ & P3, P4 via spin 11\\
        \hline
    \end{tabular}
    \caption{\footnotesize 12 spin configurations that belong to $XY$ sector, grouped by primary (P\#) and secondary (S\#) states. The column of connections indicates how primary and secondary states are related with each other. For the primary states, 4 flippable bonds are listed for each state. For the secondary states, the related primary states are listed together with the spin index necessary to flip. For example, the S1 state can be obtained from P1 state flipping the $(4-8)$ bond, or from P3 state flipping the $(5-8)$ bond. SSS rectangle is on the $xy$ plane, and the spin configuration on this plane is common for all the 12 states, as highlighted in red. The $zx$ plane supports spins staggered along the longer sides, and the spin configurations are common for all the 4 primary states, as highlighted in green.}
    \label{tab:table3}
\end{table*}

\begin{table*}[ht]
    \centering
    \begin{tabular}{|c||c|c|c|c|c|c|c|c||c|c|c|c|c|c|c|c|}
        \hline
        \rule{0pt}{1em}XY sector & 0 & 1 & 2 & 3 & $\overline{0}$ & $\overline{1}$ & $\overline{2}$ & $\overline{3}$ & 0 & 1 & 2 & 3 & $\overline{0}$ & $\overline{1}$ & $\overline{2}$ & $\overline{3}$ \\
        \hline       
        P1 & $++-$ & $-++$ & $+--$ & $--+$ & ${\color{cyan}{-}} -\color{red}{+}$ & $+--$ & $-++$ & $++-$   &   \color{red}{$-zx$} & $-$$xy$ & $+$$xy$ & $+$$zx$    & $+$$zx$ & $+$$xy$ & $-$$xy$&   $-$$zx$\\
        P2 & $++-$ & $-++$ & $+--$ & $--+$ & ${\color{cyan}{-}}+\color{red}{+}$ & $++-$ & $--+$ & $+--$   &   $-$$zx$ & $-$$xy$ & $+$$xy$ & $+$$zx$    & $-$$xy$ & $-$$zx$ & $+$$zx$  &   $+$$xy$\\
        P3 & $+--$ & $--+$ & $++-$ & $-++$ & ${\color{cyan}{-}}-\color{red}{+}$ & $+--$ & $-++$ & $++-$   &   \color{cyan}{$+$}$xy$ & $+$$zx$ & $-$$zx$ & $-$$xy$    & $+$$zx$ & $+$$xy$ & $-$$xy$ &   $-$$zx$\\
        P4 & $+--$ & $--+$ & $++-$ & $-++$ &${\color{cyan}{-}}+ \color{red}{+}$ & $++-$ & $--+$ & $+--$   &   $+$$xy$ & $+$$zx$ & $-$$zx$ & $-$$xy$    & $-$$xy$ & $-$$zx$ & $+$$zx$ &   $+$$xy$\\
        \hline
        S1 & $+-+$ & $--+$ & $+-+$ & $--+$ & ${\color{cyan}{-}}-\color{red}{+}$ & $+--$ & $-++$ & $++-$   &   $-$$yz$ & $+$$zx$ & $-$$yz$ & $+$$zx$    & $+$$zx$ & $+$$xy$ & $-$$xy$&   $-$$zx$\\
        S2 & $+-+$ & $--+$ & $+-+$ & $--+$ & ${\color{cyan}{-}}+\color{red}{+}$ & $++-$ & $--+$ & $+--$   &   $-$$yz$ & $+$$zx$ & $-$$yz$ & $+$$zx$    & $-$$xy$ & $-$$zx$ & $+$$zx$&   $+$$xy$\\
        S3 & $++-$ & $-++$ & $+--$ & $--+$ & ${\color{cyan}{-}}-\color{red}{+}$ & $+-+$ & $--+$ & $+-+$   &   $-$$zx$ & $-$$xy$ & $+$$xy$ & $+$$zx$    & $+$$zx$ & $-$$yz$ & $+$$zx$&   $-$$yz$\\
        S4 & $+--$ & $--+$ & $++-$ & $-++$ & ${\color{cyan}{-}}-\color{red}{+}$ & $+-+$ & $--+$ & $+-+$   &   $+$$xy$ & $+$$zx$ & $-$$zx$ & $-$$xy$    & $+$$zx$ & $-$$yz$ & $+$$zx$&   $-$$yz$\\
        S5 & $++-$ & $-+-$ & $++-$ & $-+-$ & ${\color{cyan}{-}}-\color{red}{+}$ & $+--$ & $-++$ & $++-$   &   $-$$zx$ & $+$$yz$ & $-$$zx$ & $+$$yz$    & $+$$zx$ & $+$$xy$ & $-$$xy$&   $-$$zx$\\
        S6 & $++-$ & $-+-$ & $++-$ & $-+-$ & ${\color{cyan}{-}}+\color{red}{+}$ & $++-$ & $--+$ & $+--$   &   $-$$zx$ & $+$$yz$ & $-$$zx$ & $+$$yz$    & $-$$xy$ & $-$$zx$ & $+$$zx$&   $+$$xy$\\
        S7 & $++-$ & $-++$ & $+--$ & $--+$ & ${\color{cyan}{-}}+\color{green}{-}$ & $++-$ & $-+-$ & $++-$   &   $-$$zx$ & $-$$xy$ & $+$$xy$ & $+$$zx$    & $+$$yz$ & $-$$zx$ & $+$$yz$&   $-$$zx$\\
        S8 & $+--$ & $--+$ & $++-$ & $-++$ & ${\color{cyan}{-}}+\color{green}{-}$ & $++-$ & $-+-$ & $++-$   &   $+$$xy$ & $+$$zx$ & $-$$zx$ & $-$$xy$    & $+$$yz$ & $-$$zx$ & $+$$yz$&   $-$$zx$\\
        \hline
    \end{tabular}
    \caption{Triplet and minority spin representations of the 12 states in the XY sector. In the triplet representation $(\sigma_i,\sigma_j,\sigma_k)$, $\sigma_i,\sigma_j$, and $\sigma_k$ are placed on $xy$, $yz$, and $zx$ planes in this order. To relate the two representations, for example, the triangle $0$ of the P1 state is $(\sigma_i,\sigma_j,\sigma_k)=(+, +, -)$, with the minority spin $\sigma_k$ is on the $zx$ plane and takes the value $-1$. Accordingly, it is expressed as $-zx$ in the minority spin representation. The selectivity of the coupling is demonstrated through a few examples.}
    \label{tab:table4}
\end{table*}

\section{Paving the lattice with icosahedra}

\subsection{Connecting triangles and spin representations}

Icosahedra are placed on a bcc lattice, which means each icosahedron interacts with eight neighbors via $J_2$ couplings connecting opposite triangles facing each other. Each icosahedron thus possesses 8 triangles interacting with its neighbors; we shall call them the connecting triangles. We label these triangles with the indices $a$ and $\overline{a}$ ($a=0,1,2,3$), so that a triangle $a$ faces with $\overline{a}$ of the neighboring icosahedron [Fig.~\ref{Fig2} (e)].
Each connecting triangle is described by a triplet of spins, $(\sigma_i,\sigma_j,\sigma_k)$, where $\sigma_i,\sigma_j$, and $\sigma_k$ are placed respectively on the $xy$, $yz$, and $zx$ planes [Table~\ref{tab:table1} (Right)].
Due to the triangle condition, each triplet $(\sigma_i,\sigma_j,\sigma_k)$ has 6 possible values in the ground state.

A trivial way to represent the spin configuration of an icosahedron is as a list of its spin orientations, as given in Table \ref{tab:table3}. This representation is, however, not ideal to determine how two neighboring icosahedra interact. The set of triplets on the 8 connecting triangles give a more convenient representation of the icosahedral ground states, which we call the triplet representation, as listed in Table~\ref{tab:table4}.

Another convenient representation is the minority spin representation, which we also list in Table~\ref{tab:table4}. Triplets can be labeled by the position and orientation of the minority spin: for example, a triplet, $(\sigma_i,\sigma_j,\sigma_k)=(++-)$ can be expressed as $-zx$, since the spin on the $zx$ plane is the only spin that takes the value, $-1$.

\subsection{Inter-icosahedral interaction}
\label{sec:Inter-icosahedral_interaction}
Each pair of connecting triangles between neighboring icosahedra is coupled via six $J_2$ couplings forming an hexagon [Fig.~\ref{Fig2} (a,f)], whose energy also needs to be minimized in order to find the ground state. Ferromagnetic $J_2$ favor the same orientation for all 6 spins. However, as long as $J_2<J_1$, the energy minimization of $J_2$ couplings shall not affect the intra-icosahedron $J_1$ interactions. Since the two connecting triangles still have to respect the triangle condition, the 6 spins cannot all be the same. The best compromise is found when 4 of the 6 ferromagnetic $J_2$ couplings are satisfied, leaving only two bonds with antiferromagnetic spin orientation. As illustrated in Fig.~\ref{Fig2}(f), it implies that if the minority spin of one of the connecting triangle is, say $\pm zx$, then the spin of the other connecting triangle belonging to the $zx$ plane will be $\sigma=\mp 1$. As long as the trinagle condition is imposed on all icosahedra, this constraint is actually the only necessary condition to minimize the $J_2$ couplings between two neighboring icosahedra.

This condition has direct consequences on the connectivity between different icosahedral sectors in the ground state of the system, that can be rationalized as follows.
As an example, let us consider one icosahedron in the P1 configuration of the XY sector [Table~\ref{tab:table4}]. In the minority spin representation, the connecting triangle $0$ is in the $-zx$ state, as highlighted in red in Table~\ref{tab:table4}.
This $-zx$ label compactly expresses the constraint on the connecting triangle $\overline{0}$ of the neighboring icosahedron [Fig.~\ref{Fig2}(f)]. As discussed just above, the energy of the ferromagnetic inter-icosahedral interaction is minimized if and only if the spin on the $zx$ plane on the connecting triangle $\overline{0}$ takes the value $+1$.

Accordingly, if the neighboring icosahedron is in the XY sector, 10 configurations highlighted in red in Table~\ref{tab:table4} are possible. In contrast, provided the neighboring icosahedron is in the $\overline{\rm XY}$ sector, all the spins are reversed. Then, only the S7 and S8 configurations are possible, as highlighted in green in Table~\ref{tab:table4}, since only these two configurations have the $zx$-spin to be $+1$ in the $\overline{\rm XY}$ sector.
It implies that one sector has more possibilities to connect to itself than to another sector. 

This tendency becomes more evident, for example, if we consider the P3 configuration of the XY sector, as highlighted in blue in Table~\ref{tab:table4}.
In this case, the connecting triangle $0$ is in the $+xy$ state, and never couples to any state of the $\overline{\rm XY}$ sector, but couples to any state of the XY sector, since the $xy$-spin is $-1$ ($+1$) for all the 12 configurations in the XY ($\overline{\rm XY}$) sector. By symmetry, this entropic preference for a given sector to connect with itself on neighboring icosahedra applies to all sectors and to all pairs of connecting triangles.

Note that here we see the impact of changing the nature of the $J_2$ coupling: it only changes the sector favored on one sublattice, going from XY for ferromagnetic $J_2$ to $\overline{\rm XY}$ for antiferromagnetic $J_2$.

This strong selectivity of the neighboring triplet can be naturally understood from the way each sector is constructed.
In each sector, the spins on the SSS plane are completely the same for the 12 configurations in the sector.
Furthermore, the spins on the LSS plane are also mostly the same; for primary states they are completely identical, while for secondary states they are half identical.
Consequently, if a certain triplet is not compatible with these fixed spins on the neighboring triangle, the coupling to the sector becomes immediately prohibited. This selectivity will be important in order to rationalize the thermal properties of our model.

\section{Order or spin liquid ?}

\subsection{Thermal properties}
\begin{figure}[ht]
\begin{center}
\includegraphics[width=0.5\textwidth]{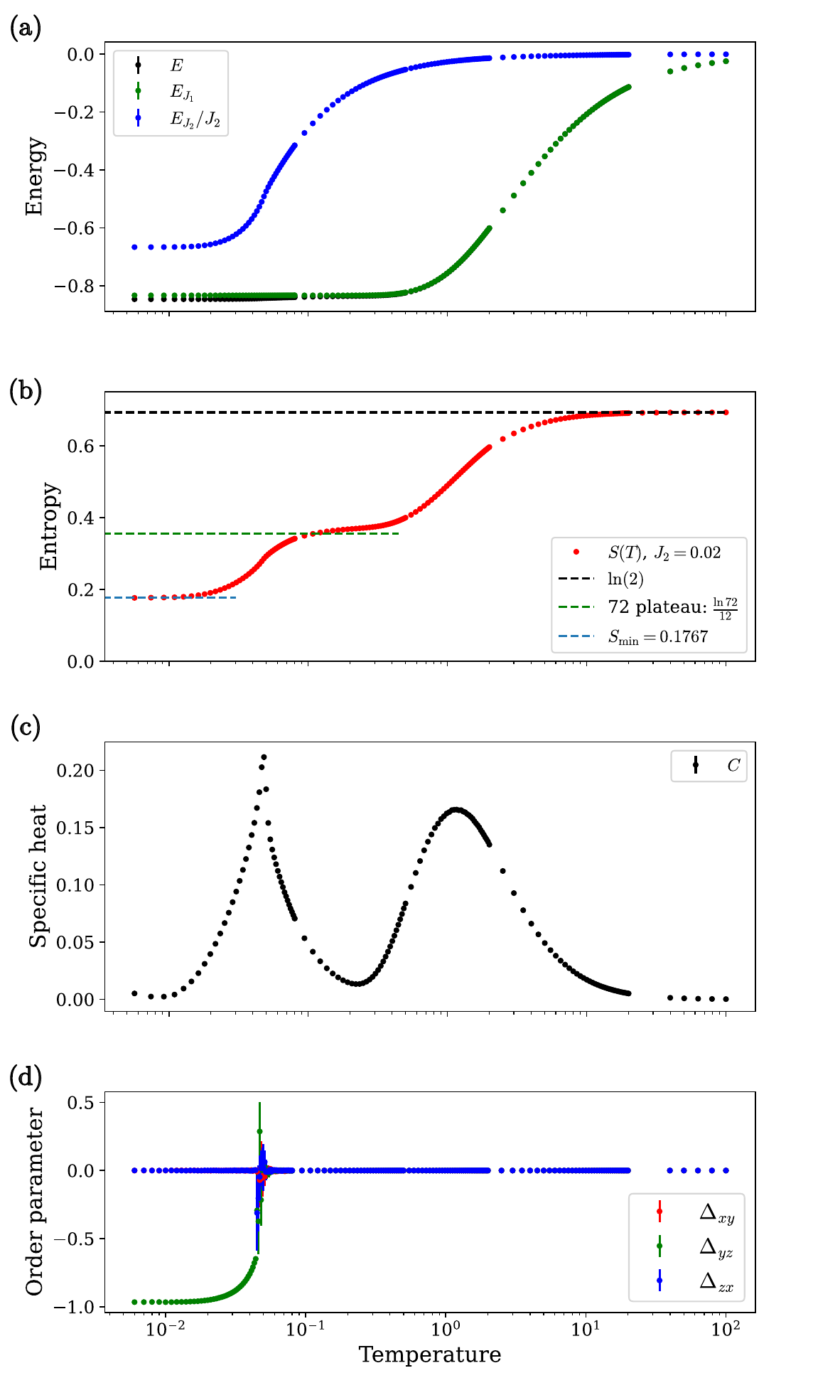}
\end{center}
\caption{\label{Fig3} 
(Color online) Temperature dependence of (a) the energy, (b) the entropy per spin, (c) the specific heat, and (d) the order parameter, for $L=14\leftrightarrow N_{\rm spin}=65856$ spins.
The values per spin are shown. In (a), in addition to the total energy, the intra- and inter-icosahedral energies are shown.}
\end{figure}

In Fig.~\ref{Fig3}, we show the temperature dependence of thermodynamic quantities, obtained by Monte Carlo simulation.
Keeping in mind that $J_1=+1$, hereafter we set $J_2=-0.02$ for the sake of convenience. Our results remain valid for a large region of the ratio $J_2/J_1$.

Firstly, in Fig.~\ref{Fig3} (c), the specific heat shows a clear double-peak structure: a broad peak appears at high temperature, $T_{\rm H}\sim 1.0$, and a sharp peak shows up at a low temperature $T_{\rm L}$.
As is naturally guessed from the energy scale of $T_{\rm H}\sim J_1$, the high-temperature broad peak is attributed to a crossover to the single-icosahedron ground states.
Below $T_{\rm H}$, each icosahedron takes one of the $72$ ground state configurations.
Accordingly, the intra-icosahedral energy $\varepsilon_1\equiv\frac{1}{N_{\rm spin}}\langle\mathcal{H}_1\rangle$ relaxes to its minimal value, $\varepsilon_1\to-\frac{5}{6}J_1$ [Fig.~\ref{Fig3} (a)].
In contrast, the inter-icosahedral energy $\varepsilon_2\equiv\frac{1}{N_{\rm spin}}\langle\mathcal{H}_2\rangle$ does not show any visible change around $T_{\rm H}$, implying that icosahedra are still mostly uncorrelated from each other in this temperature range.

\begin{figure}[ht]
\begin{center}
\includegraphics[width=0.5\textwidth]{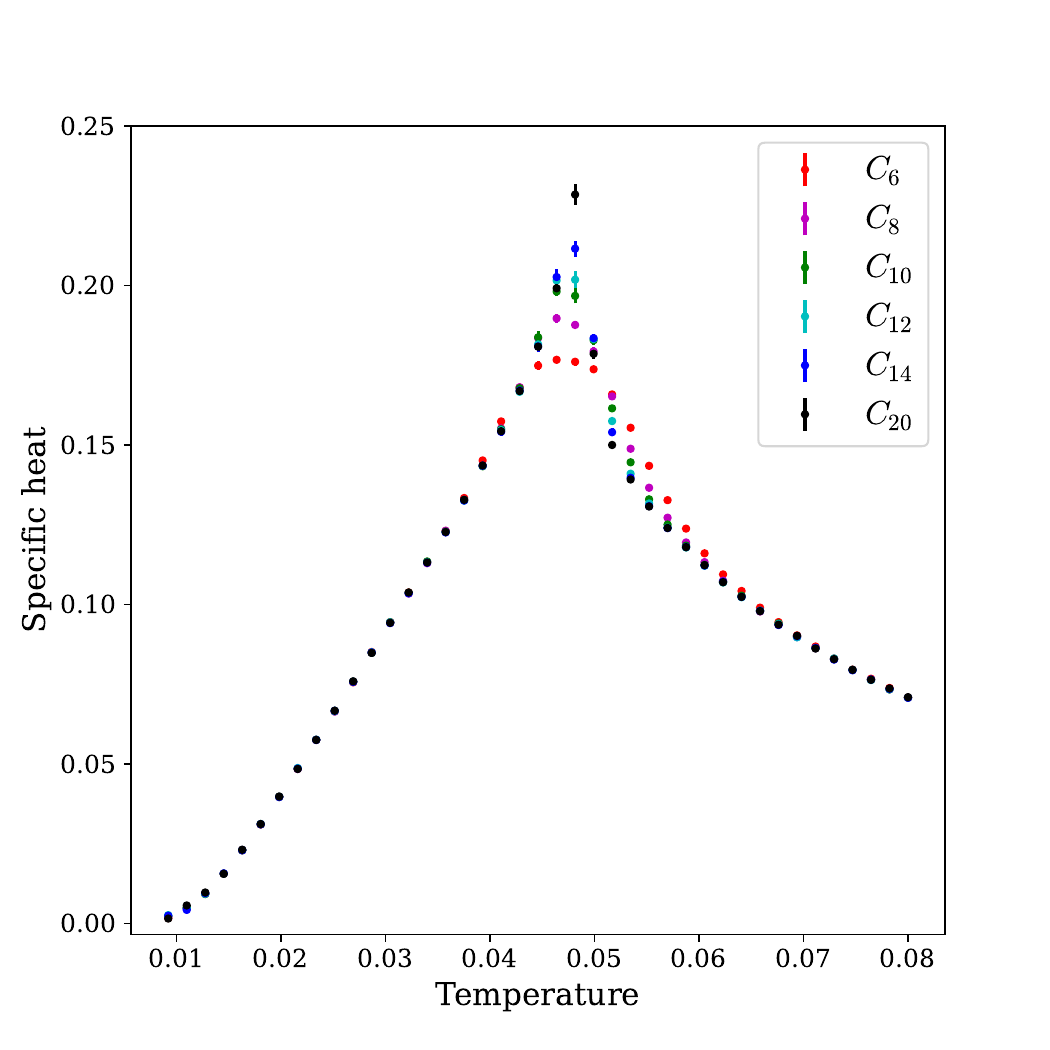}
\end{center}
\caption{\label{Fig4} 
(Color online) System size scaling of the phase transition. System size goes from $L=6\leftrightarrow N_{\rm spin}=5184$ spins to $L=20\leftrightarrow N_{\rm spin}=192000$ spins by steps of $2$. The system is well thermalised for $T\geq 10^{-2}$, but fails to thermalise for smaller temperatures. The entropy is therefore calculated down to the smallest reliable temperature for $T>10^{-2}$.}
\end{figure}

In contrast to the broad peak at $T_{\rm H}$, the low-temperature peak shows characteristic behavior of a phase transition.
In Fig.~\ref{Fig4}, the peak is shown to evolve sharply when increasing the system size, suggesting a second-order phase transition.\\

To characterize the low-temperature phase, we need to define an order parameter.
As discussed in Sec.~\ref{sec:Sector}, the degenerate ground states can be classified into 6 sectors.
To distinguish these sectors, we define the order parameter as
\begin{eqnarray}
\Delta_{\alpha} = \frac{3}{N_{\rm spin}}\sum_{{\mathbf R}_i}\sum_{j\in\alpha}(-1)^j\sigma_{{\mathbf R}_i,j},
\end{eqnarray}
for $\alpha=xy, yz,$ and $zx$. Here, ${\mathbf R}_i$ refers to the position of the icosahedron, and $j=0-11$ is the intra-icosahedral coordinate as introduced in Sec.~\ref{sec:Sector}, and listed in Table~\ref{tab:table1}, and the summation over $j$ is here limited to the plane $\alpha$.
This order parameter intends to detect SSS order on the plane $\alpha$.
Assuming SSS order is realized on the $xy$ plane, then site $0$ (resp. $1$) is ferromagnetically aligned with site $2$ (resp. $3$), so if $\sigma_0=+1 (-1)$, then $\Delta_{xy}=+1 (-1)$.
$\Delta_{\alpha}$ stays close to zero at high temperatures, but suddenly starts to rise at $T_{\rm L}$ and continues to grow following a concave curve, as is typical with continuous phase transitions.
The finite $\Delta_{\alpha}$ implies that the phase transition selects a given sector, and is associated with the $\mathbb{Z}_3\times\mathbb{Z}_2$ symmetry breaking due to the $C_{3}$ spatial rotation around $[111]$ axes, and the time-reversal operation.\\ 
 
Reflecting the double-peak structure of the specific heat, the entropy per spin shows two plateau regions below $T_{\rm L}$ and between $T_{\rm H}$ and $T_{\rm L}$, respectively.
The plateau value just below $T_{\rm H}$ is consistent with the $72$-fold degeneracy of a single icosahedron, $\mathcal{S}_{\rm ico}=\frac{\log72}{12}\approx 0.356$, giving another evidence for the peak at $T_{\rm H}$ to indicate the crossover from paramagnetism to the independent-icosahedral ground states.

We find in simulations that this first plateau between $T_{\rm H}$ and $T_{\rm L}$ persists up to $|J_2/J_1| \lesssim 0.1$. For larger values, at least up to $|J_2/J_1| \lesssim 1$, there is only one peak in the specific heat; the entropy directly reaches the second plateau and remains finite at the same value down to the lowest temperatures.

The second plateau below $T_{\rm L}$ is nontrivial. 
Naively, we would expect the entropy to drop to zero below the transition temperature, $T_{\rm L}$, where the magnetic ordering takes place, as implied by the finite value of the order parameter.
However, the entropy decrease stops at a finite value, and keeps its residual value down to zero temperature.

\subsection{Structure of the low temperature phase}
From the analysis of thermodynamic quantities presented in the previous section, we found that the low-temperature phase shows spontaneous symmetry breaking, as characterized by the order parameter, $\Delta_{\alpha}$, while keeping substantial amount of ground state entropy, $S_{\rm min}$.
How can we reconcile these apparently contradicting order and disorder aspects of the low-temperature phase ?

To gain an insight into the spin structure in the low-temperature phase, we calculate the probability distribution of the spin states in the selected sector with Monte Carlo simulation.
At the lowest temperature,
we find a marked unbalance between the primary and secondary states. Despite the fact that 2/3 of the 72 single-icosahedron ground states are secondary states, primary states have a noticeably higher probability of occurrence in simulations at low temperatures: $p_{1}\simeq0.578$ versus $p_{2}\simeq0.336$ for secondary ones. 
The rest of $1-p_{1}-p_{2}\simeq0.086$ is the probability that configurations in different sectors appear, which is consistent with the value of $|\Delta|$ being a bit smaller than $1$ at lowest temperatures.

To understand this distribution, one needs to pay attention to the inter-icosahedral connectivity.
Based on our simulations, let us assume that all icosahedra are in the same sector.
Following Sec.~\ref{sec:Inter-icosahedral_interaction} and Table \ref{tab:table4}, one primary state can link to the 12 states of the same sector on 4 connecting triangles, and to 10 states on the 4 remaining connecting triangles. 
Carrying out the same manipulation for secondary states, one finds that any secondary state can connect to 6 states of the same sector on 2 of its connecting triangles, 10 states on 4 other connecting triangles, and 12 states on the remaining 2 connecting triangles.
Accordingly, if we ignore farther connections, each primary state allows $\Omega_{\rm P}\equiv12^4\times10^4\sim2\times10^8$ possible spin configurations in the neighboring icosahedra, while a secondary state allows $\Omega_{\rm S}\equiv12^2\times10^4\times6^2=\frac{1}{4}\Omega_{\rm P}$ spin configurations on the 8 neighboring icosahedra, out of the total $\Omega_{\rm T}=12^8$ configurations, respectively. 
Considering that the number of secondary states is twice that of primary states, this ratio of possible neighboring configurations reasonably accounts for the unbalance mentioned found numerically.

\subsection{Pauling's entropy}
The connectivity of dominant and subdominant configurations presented in the previous section enables us to estimate the ground state entropy in a mean-field type argument analogous to Pauling's estimate of water ice entropy.
To this aim, we divide the $N$ icosahedra composing the bcc lattice into two sublattices, A and B.
Then, suppose that all the A sublattice is occupied by the configurations in one sector, which gives the number of the states, $12^{\frac{N}{2}}$.
Then, the configurations on A sublattice put a constraint on the possible configurations on B sublattice, which reduces its number to much less than $12^{\frac{N}{2}}$.
If the spin configuration on each of the 8 icosahedra, surrounding the B sublattice, is chosen completely randomly from the 12 states, the probability that one particular primary state to be possible on the B sublattice is $\frac{\Omega_{\rm P}}{\Omega_{\rm T}}=\Bigl(\frac{5}{6}\Bigr)^4$. We could add probabilities that allow different primary states and secondary states, but they give only small corrections.

If we estimate the constraint per icosahedron as $\frac{\Omega_{\rm P}}{\Omega_{\rm T}}$, we can estimate the residual entropy as
\begin{eqnarray}
\mathcal{S}_{\rm P}=\frac{1}{12N}\log\Bigl[12^{N}\Bigl(\frac{\Omega_{\rm P}}{\Omega_{\rm T}}\Bigr)^{\frac{N}{2}}\Bigr]\simeq0.1767/{\rm spin},
\end{eqnarray}
which is remarkably close to the numerical value, $\mathcal{S}_{\rm min}\simeq0.1769$. 

\subsection{Origin of ordering}
While the spin structure of the low-temperature phase is established, one important question still remains: Why are the spins ordered in this system ? 
Magnetic ordering usually takes place as a result of competition between energy and entropy.
This universal mechanism may be most clearly presented by Peierls's argument on the estimate of the transition temperature of the 2D Ising model.
According to the argument, starting from the perfect ferromagnetic ground state, the energy cost of creating the domain of flipped spins is estimated as $2JL$, with the length of domain boundary, $L$.
Meanwhile, the gain of configurational entropy to create the domain is estimated, e.g. by a Brownian motion argument, to be $L\log 3$. 
Balancing these effects leads to the approximate value of the transition temperature, $T_{\rm P}/J=\frac{2}{\log3}\simeq1.820$, which is reasonably close to the exact transition temperature, $T_{\rm c}/J=\frac{2}{\log(1+\sqrt{2})}\simeq2.269$.

From this viewpoint, the ordering of our model may be surprising.
Since there is extensive degeneracy in the ground state, the disturbance of the ordered configuration does not cost any energy, leading to $T_{\rm c}\to0$, at first sight.
Then, what is the driving force to align the SSS rectangles ?

The answer to this question is that the destruction of ordering, i.e. the creation of domain walls, reduces the entropy considerably, beyond the gain of configurational entropy from the domain-wall fluctuations.
In other words, the typical spin configurations in the degenerate ground state are ordered.
To see this, let us start with the ground state in one sector, and see what happens if domain boundaries are introduced by inserting a region of different sector, while keeping the system in the ground state.

To be specific, let us consider 4 icosahedra placed on the $xy$ plane, and assume they belong to the XY sector.
Then, let us look at the icosahedron on top of them, and ask how many spin configurations in the YZ sector are possible at this top icosahedron. 
Close inspection of the matching condition reveals that only 3 configurations, P2, S2, S7, turn out to be compatible, irrespective of which spin configurations in the XY sector the bottom 4 icosahedra take.
Insertion of different sectors introduces severe restriction on the connections and reduce the entropy. 
See Appendix. B for details.

This means that if the XY-YZ domain wall is generated along the $z$ direction, then, the boundary YZ icosahedron is left with the entropy of $\log 3\simeq1.58\log 2$, which is almost half of the residual entropy per icosahedron, $12\mathcal{S}_{\rm min}\simeq3.06\log 2$. This reduction of entropy cannot be compensated by the configurational entropy from the domain wall pattern.
It is not easy to evaluate the configurational entropy of a domain wall, $\mathcal{S}_{D}$, in a three-dimensional system.
However, if we estimate it from the transition temperature of 3D Ising model, $T^{\rm 3D}_{\rm c}\simeq4.512J$, we obtain $\mathcal{S}_{D}\simeq\frac{2J}{T^{\rm 3D}_{\rm c}}\simeq0.64\log2$.
Consequently, even if the configurational entropy is considered, it cannot make up for the loss of local icosahedral entropy.

This ordering mechanism is somewhat reminiscent of the so-called ``order-by-disorder" mechanism, but it differs in one respect; only entropy plays a role.
In typical order-by-disordered scenarios, the population of thermal excitations or the reduction of ground state energy by quantum fluctuation selects an ordering pattern.
In contrast to these normal stories, in this system, the selection of order is of purely entropic origin.

\section{Discussion and Summary}
We addressed the antiferromagnetic Ising model defined on the 1/1 approximant of Tsai-type quasicrystal, one of the simplest and realistic models of quasicrystal magnet with geometrical frustration.
We found the system to show a two-step change for $|J_2/J_1|\lesssim 0.1$ in thermodynamic properties as indicated by the double-peak structure of specific heat.
The higher-$T$ peak corresponds to the crossover from the high-temperature paramagnet to the intermediate icosahedral paramagnetic region, where each icosahedron relaxes to their ground states independently.
The lower-$T$ peak shows the second-order phase transition due to the combined ${\mathbb Z}_3\times{\mathbb Z}_2$ symmetry breaking.

In the low-temperature ordered phase, the ordering pattern is characterized by the orientation of the SSS rectangles on which spins take 2-up 2-down configurations with two spins aligned along the longer side.
Despite the magnetic ordering, the low-temperature phase keeps a macroscopic residual entropy. 
The coexistence of order and disorder results from the competition between the local icosahedral entropy and global configurational entropy of domain patterns.
It means that typical states of the degenerate ground state present long-range magnetic order.

The model considered in this work is one of the simplest, yet realistic models of geometrical frustration on quasicrystal magnets.
We believe that the obtained ordering pattern and its mechanism will shed a new light on the magnetic structure of the compounds reported so far, and also on future studies.
Most quasicrystal magnets have metallic tendency with finite electric conductivity, and the magnetic interaction has long-range components of RKKY type.
Nevertheless, to understand the nature of such complicated systems and models, the understanding of the simple short-range model is essential
We hope that our analysis will serve as a useful reference for further studies in this field.

In particular, the origin of ordering we have found in this system implies a general mechanism to design magnetic structure.
In this system, the ordering takes place, since the entropy loss of local icosahedron exceeds the gain of global configurational entropy of domain patterns.
This competition suggests an interesting common viewpoint in systems where local degrees of freedom are associated with a domain boundary.
In such systems, depending on the competition between local and global entropies, magnetic order may be stabilized or destabilized.
Particularly interesting is the competitive region, where both contributions are in subtle balance.
In those systems, exotic spatial magnetic structure might emerge from the keen correlation between local and global degrees of freedom.
We would like to leave this interesting question as a future problem.

We are grateful to Peter Holdsworth for giving us a precious opportunity to start this project. This work was supported by the JSPS KAKENHI (Nos. JP20H05655, JP22H01147, and JP23K22418), MEXT, Japan and by the French National Agency for Research (ANR-23-CE30-0038-01 and ANR-25-CE30-5029-04) and Idex Bordeaux (Research Program GPR Light).

\appendix
\section{Construction of single icosahedral ground state}
\label{Appendix:GSsingleicosahedron}
In the single icosahedral limit $J_2=0$, the ground state has 72 configurations.
To find all the configurations, let us define the effective potential at vertex $j$ as $V_j\equiv-\sum'_{j'}\sigma_{j'}$. 
Here, the summation of $j'$ is taken over the nearest-neighbor sites of $j$.
Since each site $j$ has 5 neighbors, the effective potential takes 6 possible values, $V_j=-5, -3, -1, 1, 3, 5$. The total energy of the icosahedron is expressed as
\begin{eqnarray}
E = -\frac{1}{2}\sum_{j=0}^{11}\sigma_jV_j.
\end{eqnarray}
\begin{figure}[ht]
    \begin{center}
        \includegraphics[width=\linewidth]{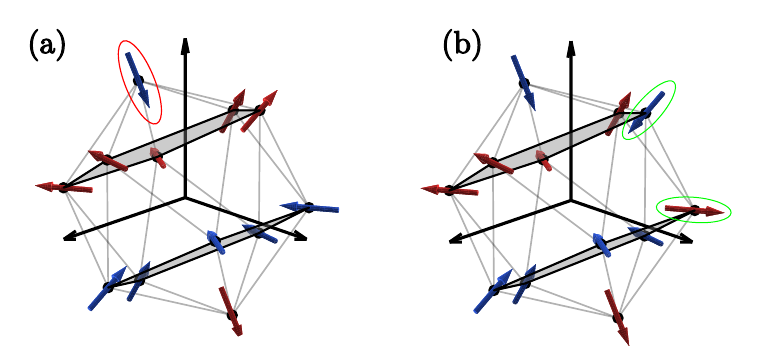}
    \end{center}
    \caption{\label{Fig5} (a) Single icosahedral ground state with the reference spin circled in red having $V_j=5$. The other spins are fully constrained by the triangle rule. (b) Single icosahedral ground state with the reference spin having $V_j=3$. Note that it is connected to (a) by flipping the two spins circled in green (see Sec.~\ref{sec:Sector}).}
    \label{fig:placeholder}
\end{figure}
Meanwhile, since all the 20 triangles contributes to the energy $-1$ in the ground state, the ground state energy is equal to $-10$, then it leads to
\begin{eqnarray}
\sum_{j=0}^{11}\sigma_jV_j = 20.
\label{eq:effV1}
\end{eqnarray}
Now, we consider the maximal value of $V_j$, $V_{\rm max}$, for a ground-state spin configuration.
If $V_{\rm max}\leq1$,
\begin{eqnarray}
\sum_{j=0}^{11}\sigma_jV_j \leq12,
\end{eqnarray}
and cannot satisfy Eq.~(\ref{eq:effV1}). So, $V_{\rm max}=3$ or $5$.
First, suppose $V_{\rm max}=5$, and set $V_j=5$ for a site $j$.
If we assume $\sigma_j=+1$, the site $j$ is surrounded by five sites with $\sigma_{j'}=-1$.
Then, the triangle condition determines the values of the rest of spins.
The number of spin configurations of this type is 12, according to the choice of the site $j$ on the icosahedron.

Next, suppose $V_{\rm max}=3$, and set $V_j=3$ for a site $j$, and assume $\sigma_j=+1$, then the site $j$ is surrounded by four sites with $\sigma_{j'}=-1$, and one site with $\sigma_{j'}=+1$.
Again, the triangle condition determines the values of the rest of spins, except for the choice of the two sites with $\sigma=+1$.
Counting the possible number of spin configurations constructed this way, we find $12\times 5= 60$ configurations of this type.

Please note that this counting argument takes into account time reversal symmetry (by symmetry of the ground states, see Fig.~\ref{Fig5}) and results in a total of $60+12=72$ ground states.

\section{Connection between XY and YZ sectors}
To see the entropy reduction by creating a domain wall between the XY and YZ sectors, we compare the triplet and spin minority representations of the both sectors [Table~\ref{tab:tableApp1}].
Suppose that we place an icosahedron in the YZ sector on the surface of XY ordered domain.
First, we assume that the surface in the $xy$ plane, i.e. the same as the SSS plane of the XY sector.

For the connection to be possible, the minority spin on the YZ icosahedron must be compatible with the spins on the SSS plane of the XY icosahedra.
The color of red and green must match in Table~\ref{tab:tableApp1}.
As an example of prohibited connection, the P1 state of the YZ sector takes $+xy$ triplet state on the triangle $\overline{0}$, which is incompatible with the minority spin $+1$ on the $xy$ plane of the triangle $0$ of the XY icosahedron.
Similarly, there might occur conflicts for the $0-\overline{0}$ and $3-\overline{3}$ links: P1, P3, S1, S3, S4, and S5 states are incompatible with the minority spin $+1$ on the $0-\overline{0}$ link, and P3, P4, S4, S5, S6, and S8 states are incompatible with the minority spin $-1$ on the $3-\overline{3}$ link.
As a result, only the three states, P3, S4 and S5 can be connected to the basal 4 icosahedra in the XY sector.

Domain wall creation on the $yz$ plane is even worse.
In this case, we need to consider the connection between the triangles $0, 1, \overline{2}, \overline{3}$ in the XY sector, and  $\overline{0}, \overline{1}, 2, 3$ in the YZ sector.
Checking the color matching as above, we found only two states, S5 and S6 survive, leading to larger entropy reduction.

\begin{table*}[ht]
    \centering
    \begin{tabular}{|c||c|c|c|c|c|c|c|c||c|c|c|c|c|c|c|c|}
        \hline
        \rule{0pt}{1em}XY sector & 0 & 1 & 2 & 3 & $\overline{0}$ & $\overline{1}$ & $\overline{2}$ & $\overline{3}$ & 0 & 1 & 2 & 3 & $\overline{0}$ & $\overline{1}$ & $\overline{2}$ & $\overline{3}$ \\
        \hline       
        P1 & ${\color{green}{+}}+-$ & ${\color{red}{-}}++$ & ${\color{green}{+}}--$ & ${\color{red}{-}}-+$ & ${\color{red}{-}}-+$ & ${\color{green}{+}}--$ & ${\color{red}{-}}++$ & ${\color{green}{+}}+-$   &   $-zx$ & $-$$xy$ & $+$$xy$ & $+$$zx$    & $+$$zx$ & $+$$xy$ & $-$$xy$&   $-$$zx$\\
        P2 & ${\color{green}{+}}+-$ & ${\color{red}{-}}++$ & ${\color{green}{+}}--$ & ${\color{red}{-}}-+$ & ${\color{red}{-}}++$ & ${\color{green}{+}}+-$ & ${\color{red}{-}}-+$ & ${\color{green}{+}}--$   &   $-$$zx$ & $-$$xy$ & $+$$xy$ & $+$$zx$    & $-$$xy$ & $-$$zx$ & $+$$zx$  &   $+$$xy$\\
        P3 & ${\color{green}{+}}--$ & ${\color{red}{-}}-+$ & ${\color{green}{+}}+-$ & ${\color{red}{-}}++$ & ${\color{red}{-}}-+$ & ${\color{green}{+}}--$ & ${\color{red}{-}}++$ & ${\color{green}{+}}+-$   &   $+$$xy$ & $+$$zx$ & $-$$zx$ & $-$$xy$    & $+$$zx$ & $+$$xy$ & $-$$xy$ &   $-$$zx$\\
        P4 & ${\color{green}{+}}--$ & ${\color{red}{-}}-+$ & ${\color{green}{+}}+-$ & ${\color{red}{-}}++$ &${\color{red}{-}}++$ & ${\color{green}{+}}+-$ & ${\color{red}{-}}-+$ & ${\color{green}{+}}--$   &   $+$$xy$ & $+$$zx$ & $-$$zx$ & $-$$xy$    & $-$$xy$ & $-$$zx$ & $+$$zx$ &   $+$$xy$\\
        \hline
        S1 & ${\color{green}{+}}-+$ & ${\color{red}{-}}-+$ & ${\color{green}{+}}-+$ & ${\color{red}{-}}-+$ & ${\color{red}{-}}-+$ & ${\color{green}{+}}--$ & ${\color{red}{-}}++$ & ${\color{green}{+}}+-$   &   $-$$yz$ & $+$$zx$ & $-$$yz$ & $+$$zx$    & $+$$zx$ & $+$$xy$ & $-$$xy$&   $-$$zx$\\
        S2 & ${\color{green}{+}}-+$ & ${\color{red}{-}}-+$ & ${\color{green}{+}}-+$ & ${\color{red}{-}}-+$ & ${\color{red}{-}}++$ & ${\color{green}{+}}+-$ & ${\color{red}{-}}-+$ & ${\color{green}{+}}--$   &   $-$$yz$ & $+$$zx$ & $-$$yz$ & $+$$zx$    & $-$$xy$ & $-$$zx$ & $+$$zx$&   $+$$xy$\\
        S3 & ${\color{green}{+}}+-$ & ${\color{red}{-}}++$ & ${\color{green}{+}}--$ & ${\color{red}{-}}-+$ & ${\color{red}{-}}-+$ & ${\color{green}{+}}-+$ & ${\color{red}{-}}-+$ & ${\color{green}{+}}-+$   &   $-$$zx$ & $-$$xy$ & $+$$xy$ & $+$$zx$    & $+$$zx$ & $-$$yz$ & $+$$zx$&   $-$$yz$\\
        S4 & ${\color{green}{+}}--$ & ${\color{red}{-}}-+$ & ${\color{green}{+}}+-$ & ${\color{red}{-}}++$ & ${\color{red}{-}}-+$ & ${\color{green}{+}}-+$ & ${\color{red}{-}}-+$ & ${\color{green}{+}}-+$   &   $+$$xy$ & $+$$zx$ & $-$$zx$ & $-$$xy$    & $+$$zx$ & $-$$yz$ & $+$$zx$&   $-$$yz$\\
        S5 & ${\color{green}{+}}+-$ & ${\color{red}{-}}+-$ & ${\color{green}{+}}+-$ & ${\color{red}{-}}+-$ & ${\color{red}{-}}-+$ & ${\color{green}{+}}--$ & ${\color{red}{-}}++$ & ${\color{green}{+}}+-$   &   $-$$zx$ & $+$$yz$ & $-$$zx$ & $+$$yz$    & $+$$zx$ & $+$$xy$ & $-$$xy$&   $-$$zx$\\
        S6 & ${\color{green}{+}}+-$ & ${\color{red}{-}}+-$ & ${\color{green}{+}}+-$ & ${\color{red}{-}}+-$ & ${\color{red}{-}}++$ & ${\color{green}{+}}+-$ & ${\color{red}{-}}-+$ & ${\color{green}{+}}--$   &   $-$$zx$ & $+$$yz$ & $-$$zx$ & $+$$yz$    & $-$$xy$ & $-$$zx$ & $+$$zx$&   $+$$xy$\\
        S7 & ${\color{green}{+}}+-$ & ${\color{red}{-}}++$ & ${\color{green}{+}}--$ & ${\color{red}{-}}-+$ & ${\color{red}{-}}+-$ & ${\color{green}{+}}+-$ & ${\color{red}{-}}+-$ & ${\color{green}{+}}+-$   &   $-$$zx$ & $-$$xy$ & $+$$xy$ & $+$$zx$    & $+$$yz$ & $-$$zx$ & $+$$yz$&   $-$$zx$\\
        S8 & ${\color{green}{+}}--$ & ${\color{red}{-}}-+$ & ${\color{green}{+}}+-$ & ${\color{red}{-}}++$ & ${\color{red}{-}}+-$ & ${\color{green}{+}}+-$ & ${\color{red}{-}}+-$ & ${\color{green}{+}}+-$   &   $+$$xy$ & $+$$zx$ & $-$$zx$ & $-$$xy$    & $+$$yz$ & $-$$zx$ & $+$$yz$&   $-$$zx$\\
        \hline
    \end{tabular}
     \begin{tabular}{|c||c|c|c|c|c|c|c|c||c|c|c|c|c|c|c|c|}
        \hline
        \rule{0pt}{1em}YZ sector & 0 & 1 & 2 & 3 & $\overline{0}$ & $\overline{1}$ & $\overline{2}$ & $\overline{3}$ & 0 & 1 & 2 & 3 & $\overline{0}$ & $\overline{1}$ & $\overline{2}$ & $\overline{3}$ \\
        \hline       
        P1 & $-++$ & $-++$ & $+-+$ & $+-+$       & $+--$ & $+--$ & $-+-$ & $-+-$   &  \color{green}{$-$$xy$} & \color{green}{$-$$xy$} & $-$$yz$ & $-$$yz$    & \color{red}{$+$$xy$} & \color{red}{$+$$xy$} & $+$$yz$ &  $+$$yz$\\
        P2 & $-++$ & $-+-$ & $+-+$ & $+--$        & $+-+$ & $+--$ & $-++$ & $-+-$   &   \color{green}{$-$$xy$} & $+$$yz$ & $-$$yz$ & \color{red}{$+$$xy$}     & $-$$yz$ & \color{red}{$+$$xy$} & \color{green}{$-$$xy$} &   $+$$yz$\\
        P3 & $-+-$ & $-++$ & $+--$ & $+-+$       & $+--$ & $+-+$ & $-+-$ & $-++$   &   $+$$yz$ & \color{green}{$-$$xy$} & \color{red}{$+$$xy$} & $-$$yz$    & \color{red}{$+$$xy$} & $-$$yz$ & $+$$yz$ &   \color{green}{$-$$xy$}\\
        P4 & $-+-$ & $-+-$ & $+--$ & $+--$       &$+-+$ & $+-+$ & $-++$ & $-++$   &   $+$$yz$ & $+$$yz$ & \color{red}{$+$$xy$} & \color{red}{$+$$xy$}    & $-$$yz$ & $-$$yz$ & \color{green}{$-$$xy$} &   \color{green}{$-$$xy$}\\
        \hline
        S1 & $++-$ & $-++$ & $+--$ & $+-+$       & $+--$ & $+--$ & $-+-$ & $++-$   &   $-$$zx$ &   \color{green}{$-$$xy$} & \color{red}{$+$$xy$} & $-$$yz$     & \color{red}{$+$$xy$} & \color{red}{$+$$xy$} & $+$$yz$& $-$$zx$\\
        S2 & $++-$ & $-+-$ & $+--$ & $+--$       & $+-+$ & $+--$ & $-++$ & $++-$   &   $-$$zx$ &   $+$$yz$ & \color{red}{$+$$xy$} & \color{red}{$+$$xy$}    & $-$$yz$ & \color{red}{$+$$xy$} & \color{green}{$-$$xy$} & $-$$zx$\\
        S3 & $-++$ & $++-$ & $+-+$ & $+--$       & $+--$ & $+--$ & $++-$ & $-+-$   &   \color{green}{$-$$xy$} &   $-$$zx$ & $-$$yz$ & \color{red}{$+$$xy$}    & \color{red}{$+$$xy$} & \color{red}{$+$$xy$} & $-$$zx$ & $+$$yz$\\
        S4 & $-+-$ & $++-$ & $+--$ & $+--$       & $+--$ & $+-+$ & $++-$ & $-++$   &   $+$$yz$  &   $-$$zx$ & \color{red}{$+$$xy$} & \color{red}{$+$$xy$}   & \color{red}{$+$$xy$} & $-$$yz$ & $-$$zx$ & \color{green}{$-$$xy$}\\
        S5 & $-++$ & $-++$ & $--+$ & $+-+$       & $+--$ & $--+$ & $-+-$ & $-++$   &   \color{green}{$-$$xy$} &   \color{green}{$-$$xy$} & $+$$zx$ & $-$$yz$    & \color{red}{$+$$xy$} & $+$$zx$ & $+$$yz$ & \color{green}{$-$$xy$}\\
        S6 & $-++$ & $-+-$ & $--+$ & $+--$       & $+-+$ & $--+$ & $-++$ & $-++$   &   \color{green}{$-$$xy$} &   $+$$yz$ & $+$$zx$ & \color{red}{$+$$xy$}    & $-$$yz$ & $+$$zx$ & \color{green}{$-$$xy$} & \color{green}{$-$$xy$}\\
        S7 & $-++$ & $-++$ & $+-+$ & $--+$       & $--+$ & $+--$ & $-++$ & $-+-$   &   \color{green}{$-$$xy$} &   \color{green}{$-$$xy$} & $-$$yz$ & $+$$zx$    & $+$$zx$ & \color{red}{$+$$xy$} & \color{green}{$-$$xy$} & $+$$yz$\\
        S8 & $-+-$ & $-++$ & $+--$ & $--+$       & $--+$ & $+-+$ & $-++$ & $-++$   &   $+$$yz$ &   \color{green}{$-$$xy$} & \color{red}{$+$$xy$} & $+$$zx$    & $+$$zx$ & $-$$yz$ & \color{green}{$-$$xy$} & \color{green}{$-$$xy$}\\
        \hline
    \end{tabular}
    \caption{The triplet and minority spin representations of (top) XY and (bottom) YZ sectors are shown. In the columns of triplet representation of the top table, the spins on the SSS plane is highlighted, with $+1 (-1)$ in green (red).
    In the columns of minority spin representation of the bottom table, the $+xy (-xy)$ is highlighted in red (green).}
    \label{tab:tableApp1}
\end{table*}

\bibliographystyle{apsrev4-1}
\bibliography{IcosahedralApproximant}

@article{doi:10.7566/JPSJ.93.045001,
	author = {Sugimoto ,Takanori and Suzuki ,Shintaro and Tamura ,Ryuji and Tohyama ,Takami},
	doi = {10.7566/JPSJ.93.045001},
	eprint = {https://doi.org/10.7566/JPSJ.93.045001},
	journal = {Journal of the Physical Society of Japan},
	number = {4},
	pages = {045001},
	title = {Revisiting the Magnetic Orders in Heisenberg Model for Tsai-Type Quasicrystal Approximant},
	url = {https://doi.org/10.7566/JPSJ.93.045001},
	volume = {93},
	year = {2024}}

@article{4zb8-2zjq,
	author = {Watanabe, Shinji and Yamada, Tsunetomo and Takakura, Hiroyuki and Fujita, Nobuhisa},
	doi = {10.1103/4zb8-2zjq},
	issue = {4},
	journal = {Phys. Rev. Res.},
	month = {Oct},
	numpages = {14},
	pages = {043113},
	publisher = {American Physical Society},
	title = {Monte Carlo study on critical exponents of the classical Heisenberg model in ferromagnetic icosahedral quasicrystal},
	url = {https://link.aps.org/doi/10.1103/4zb8-2zjq},
	volume = {7},
	year = {2025}}

@article{doi:10.7566/JPSJ.90.063701,
	author = {Watanabe ,Shinji and Kawamoto ,Mina},
	doi = {10.7566/JPSJ.90.063701},
	eprint = {https://doi.org/10.7566/JPSJ.90.063701},
	journal = {Journal of the Physical Society of Japan},
	number = {6},
	pages = {063701},
	title = {Crystalline Electronic Field in Rare-Earth Based Quasicrystal and Approximant: Analysis of Quantum Critical Au--Al--Yb Quasicrystal and Approximant},
	url = {https://doi.org/10.7566/JPSJ.90.063701},
	volume = {90},
	year = {2021}}

@article{doi:10.7566/JPSJ.95.044705,
	author = {Watanabe ,Shinji and Iwasaki ,Tatsuya},
	doi = {10.7566/JPSJ.95.044705},
	eprint = {https://doi.org/10.7566/JPSJ.95.044705},
	journal = {Journal of the Physical Society of Japan},
	number = {4},
	pages = {044705},
	title = {Non-Collinear and Non-Coplanar Magnetic Orders in 1/1 Periodic Approximant to the Icosahedral Quasicrystal},
	url = {https://doi.org/10.7566/JPSJ.95.044705},
	volume = {95},
	year = {2026}}

@article{PhysRevB.110.195117,
	author = {Yan, Han and Romh\'anyi, Judit and Thomasen, Andreas and Shannon, Nic},
	doi = {10.1103/PhysRevB.110.195117},
	issue = {19},
	journal = {Phys. Rev. B},
	month = {Nov},
	numpages = {8},
	pages = {195117},
	publisher = {American Physical Society},
	title = {Pinch points and half moons encode Berry curvature},
	url = {https://link.aps.org/doi/10.1103/PhysRevB.110.195117},
	volume = {110},
	year = {2024}}

@article{PhysRevB.109.174421,
	author = {Yan, Han and Benton, Owen and Nevidomskyy, Andriy H. and Moessner, Roderich},
	doi = {10.1103/PhysRevB.109.174421},
	issue = {17},
	journal = {Phys. Rev. B},
	month = {May},
	numpages = {35},
	pages = {174421},
	publisher = {American Physical Society},
	title = {Classification of classical spin liquids: Detailed formalism and suite of examples},
	url = {https://link.aps.org/doi/10.1103/PhysRevB.109.174421},
	volume = {109},
	year = {2024}}

@article{kermarrec2025magnetizationmagnetostrictionmeasurementsdipoleoctupole,
	author = {Edwin Kermarrec and Guanyue Chen and Hiromu Okamoto and Chun-Jiong Huang and Han Yan and Jian Yan and Hikaru Takeda and Yusei Shimizu and Evan M. Smith and Avner Fitterman and Andrea D. Bianchi and Bruce D. Gaulin and Minoru Yamashita},
	journal = {arXiv:2509.09189},
	title = {Magnetization and magnetostriction measurements of the dipole-octupole quantum spin ice candidate Ce2Hf2O7},
	year = {2025}}

@article{PhysRevLett.110.217209,
	author = {Toews, W. H. and Zhang, Songtian S. and Ross, K. A. and Dabkowska, H. A. and Gaulin, B. D. and Hill, R. W.},
	doi = {10.1103/PhysRevLett.110.217209},
	issue = {21},
	journal = {Phys. Rev. Lett.},
	month = {May},
	numpages = {5},
	pages = {217209},
	publisher = {American Physical Society},
	title = {Thermal Conductivity of ${\mathrm{Ho}}_{2}{\mathrm{Ti}}_{2}{\mathbf{O}}_{7}$ along the [111] Direction},
	url = {https://link.aps.org/doi/10.1103/PhysRevLett.110.217209},
	volume = {110},
	year = {2013}}

@article{PhysRevB.88.054406,
	author = {Kolland, G. and Valldor, M. and Hiertz, M. and Frielingsdorf, J. and Lorenz, T.},
	doi = {10.1103/PhysRevB.88.054406},
	issue = {5},
	journal = {Phys. Rev. B},
	month = {Aug},
	numpages = {8},
	pages = {054406},
	publisher = {American Physical Society},
	title = {Anisotropic heat transport via monopoles in the spin-ice compound Dy${}_{2}$Ti${}_{2}$O${}_{7}$},
	url = {https://link.aps.org/doi/10.1103/PhysRevB.88.054406},
	volume = {88},
	year = {2013}}

@article{PhysRevB.86.060402,
	author = {Kolland, G. and Breunig, O. and Valldor, M. and Hiertz, M. and Frielingsdorf, J. and Lorenz, T.},
	doi = {10.1103/PhysRevB.86.060402},
	issue = {6},
	journal = {Phys. Rev. B},
	month = {Aug},
	numpages = {4},
	pages = {060402},
	publisher = {American Physical Society},
	title = {Thermal conductivity and specific heat of the spin-ice compound Dy${}_{2}$Ti${}_{2}$O${}_{7}$: Experimental evidence for monopole heat transport},
	url = {https://link.aps.org/doi/10.1103/PhysRevB.86.060402},
	volume = {86},
	year = {2012}}

@article{PhysRevB.105.104405,
	author = {Sutcliffe, Ruairidh and Rau, Jeffrey G.},
	doi = {10.1103/PhysRevB.105.104405},
	issue = {10},
	journal = {Phys. Rev. B},
	month = {Mar},
	numpages = {19},
	pages = {104405},
	publisher = {American Physical Society},
	title = {Thermal conductivity of square ice},
	url = {https://link.aps.org/doi/10.1103/PhysRevB.105.104405},
	volume = {105},
	year = {2022}}

@article{Snyder:2001aa,
	author = {Snyder, J. and Slusky, J. S. and Cava, R. J. and Schiffer, P.},
	date = {2001/09/01},
	doi = {10.1038/35092516},
	id = {Snyder2001},
	isbn = {1476-4687},
	journal = {Nature},
	number = {6851},
	pages = {48--51},
	title = {How `spin ice'freezes},
	url = {https://doi.org/10.1038/35092516},
	volume = {413},
	year = {2001}}

@article{Matsuhira:2001aa,
	author = {K Matsuhira and Y Hinatsu and T Sakakibara},
	date = {2001/07/19},
	doi = {10.1088/0953-8984/13/31/101},
	isbn = {0953-8984},
	journal = {Journal of Physics: Condensed Matter},
	number = {31},
	pages = {L737},
	title = {Novel dynamical magnetic properties in the spin ice compound Dy2Ti2O7},
	url = {https://doi.org/10.1088/0953-8984/13/31/101},
	volume = {13},
	year = {2001}}

@article{Matsuhira:2000aa,
	author = {K Matsuhira and Y Hinatsu and K Tenya and T Sakakibara},
	date = {2000/10/09},
	doi = {10.1088/0953-8984/12/40/103},
	isbn = {0953-8984},
	journal = {Journal of Physics: Condensed Matter},
	number = {40},
	pages = {L649},
	title = {Low temperature magnetic properties of frustrated pyrochlore ferromagnets Ho2Sn2O7 and Ho2Ti2O7},
	url = {https://doi.org/10.1088/0953-8984/12/40/103},
	volume = {12},
	year = {2000}}

@article{Uehara:2022aa,
	author = {Uehara, Taiki and Ohtsuki, Takumi and Udagawa, Masafumi and Nakatsuji, Satoru and Machida, Yo},
	date = {2022/08/06},
	doi = {10.1038/s41467-022-32375-0},
	id = {Uehara2022},
	isbn = {2041-1723},
	journal = {Nature Communications},
	number = {1},
	pages = {4604},
	title = {Phonon thermal Hall effect in a metallic spin ice},
	url = {https://doi.org/10.1038/s41467-022-32375-0},
	volume = {13},
	year = {2022}}

@article{Tang:2023aa,
	author = {Tang, Nan and Gritsenko, Yulia and Kimura, Kenta and Bhattacharjee, Subhro and Sakai, Akito and Fu, Mingxuan and Takeda, Hikaru and Man, Huiyuan and Sugawara, Kento and Matsumoto, Yosuke and Shimura, Yasuyuki and Wen, Jiajia and Broholm, Collin and Sawa, Hiroshi and Takigawa, Masashi and Sakakibara, Toshiro and Zherlitsyn, Sergei and Wosnitza, Joachim and Moessner, Roderich and Nakatsuji, Satoru},
	date = {2023/01/01},
	doi = {10.1038/s41567-022-01816-4},
	id = {Tang2023},
	isbn = {1745-2481},
	journal = {Nature Physics},
	number = {1},
	pages = {92--98},
	title = {Spin--orbital liquid state and liquid--gas metamagnetic transition on a pyrochlore lattice},
	url = {https://doi.org/10.1038/s41567-022-01816-4},
	volume = {19},
	year = {2023}}

@article{Matsuhira:2002aa,
	author = {K Matsuhira and Z Hiroi and T Tayama and S Takagi and T Sakakibara},
	date = {2002/07/11},
	doi = {10.1088/0953-8984/14/29/101},
	isbn = {0953-8984},
	journal = {Journal of Physics: Condensed Matter},
	number = {29},
	pages = {L559},
	title = {A new macroscopically degenerate ground state in the spin ice compound Dy2Ti2O7 under a magnetic field},
	url = {https://dx.doi.org/10.1088/0953-8984/14/29/101},
	volume = {14},
	year = {2002}}

@article{Poree:2025aa,
	author = {Por{\'e}e, Victor and Yan, Han and Desrochers, F{\'e}lix and Petit, Sylvain and Lhotel, Elsa and Appel, Markus and Ollivier, Jacques and Kim, Yong Baek and Nevidomskyy, Andriy H. and Sibille, Romain},
	date = {2025/01/01},
	doi = {10.1038/s41567-024-02711-w},
	id = {Por{\'e}e2025},
	isbn = {1745-2481},
	journal = {Nature Physics},
	number = {1},
	pages = {83--88},
	title = {Evidence for fractional matter coupled to an emergent gauge field in a quantum spin ice},
	url = {https://doi.org/10.1038/s41567-024-02711-w},
	volume = {21},
	year = {2025}}

@article{tang2025observationspinseebeckeffect,
	author = {Nan Tang and Stephan Glamsch and Aisha Aqeel and Ludwig Scheuchenpflug and Michael Schulze and Christoph Liebald and Daniel Rytz and Christo Guguschev and Manfred Albrecht and Philipp Gegenwart},
	journal = {arXiv:2509.18422},
	pages = {https://arxiv.org/abs/2509.18422},
	title = {Observation via spin Seebeck effect of macroscopic magnetic transport from emergent magnetic monopoles},
	year = {2025}}

@article{PhysRevB.96.085136,
	author = {Chen, Gang},
	doi = {10.1103/PhysRevB.96.085136},
	issue = {8},
	journal = {Phys. Rev. B},
	month = {Aug},
	numpages = {6},
	pages = {085136},
	publisher = {American Physical Society},
	title = {Spectral periodicity of the spinon continuum in quantum spin ice},
	url = {https://link.aps.org/doi/10.1103/PhysRevB.96.085136},
	volume = {96},
	year = {2017}}

@article{doi:10.1126/science.1177582,
	author = {T. Fennell and P. P. Deen and A. R. Wildes and K. Schmalzl and D. Prabhakaran and A. T. Boothroyd and R. J. Aldus and D. F. McMorrow and S. T. Bramwell},
	doi = {10.1126/science.1177582},
	eprint = {https://www.science.org/doi/pdf/10.1126/science.1177582},
	journal = {Science},
	number = {5951},
	pages = {415-417},
	title = {Magnetic Coulomb Phase in the Spin Ice Ho<sub>2</sub>Ti<sub>2</sub>O<sub>7</sub>},
	url = {https://www.science.org/doi/abs/10.1126/science.1177582},
	volume = {326},
	year = {2009}}

@article{ShintaroSuzuki2021MT-MB2020014,
	author = {Shintaro Suzuki and Asuka Ishikawa and Tsunetomo Yamada and Takanori Sugimoto and Akira Sakurai and Ryuji Tamura},
	journal = {MATERIALS TRANSACTIONS},
	number = {3},
	pages = {298-306},
	title = {Magnetism of Tsai-Type Quasicrystal Approximants},
	volume = {62},
	year = {2021}}

@article{PhysRevLett.70.3339,
	author = {Sachdev, Subir and Ye, Jinwu},
	doi = {10.1103/PhysRevLett.70.3339},
	issue = {21},
	journal = {Phys. Rev. Lett.},
	month = {May},
	numpages = {0},
	pages = {3339--3342},
	publisher = {American Physical Society},
	title = {Gapless spin-fluid ground state in a random quantum Heisenberg magnet},
	url = {https://link.aps.org/doi/10.1103/PhysRevLett.70.3339},
	volume = {70},
	year = {1993}}

@article{PhysRevLett.86.1881,
	author = {Moessner, R. and Sondhi, S. L.},
	doi = {10.1103/PhysRevLett.86.1881},
	issue = {9},
	journal = {Phys. Rev. Lett.},
	month = {Feb},
	numpages = {0},
	pages = {1881--1884},
	publisher = {American Physical Society},
	title = {Resonating Valence Bond Phase in the Triangular Lattice Quantum Dimer Model},
	url = {https://link.aps.org/doi/10.1103/PhysRevLett.86.1881},
	volume = {86},
	year = {2001}}

@article{Khomskii:2012aa,
	author = {Khomskii, D. I.},
	date = {2012/06/19},
	doi = {10.1038/ncomms1904},
	id = {Khomskii2012},
	isbn = {2041-1723},
	journal = {Nature Communications},
	number = {1},
	pages = {904},
	title = {Electric dipoles on magnetic monopoles in spin ice},
	url = {https://doi.org/10.1038/ncomms1904},
	volume = {3},
	year = {2012}}

@article{Khomskii:2021aa,
	author = {Khomskii, D. I.},
	date = {2021/05/24},
	doi = {10.1038/s41467-021-23380-w},
	id = {Khomskii2021},
	isbn = {2041-1723},
	journal = {Nature Communications},
	number = {1},
	pages = {3047},
	title = {Electric activity at magnetic moment fragmentation in spin ice},
	url = {https://doi.org/10.1038/s41467-021-23380-w},
	volume = {12},
	year = {2021}}

@article{doi:10.7566/JPSJ.90.123705,
	author = {Takatsu ,Hiroshi and Goto ,Kazuki and Sato ,Taku J. and Lynn ,Jeffrey W. and Matsubayashi ,Kazuyuki and Uwatoko ,Yoshiya and Higashinaka ,Ryuji and Matsuhira ,Kazuyuki and Hiroi ,Zenji and Kadowaki ,Hiroaki},
	doi = {10.7566/JPSJ.90.123705},
	eprint = {https://doi.org/10.7566/JPSJ.90.123705},
	journal = {Journal of the Physical Society of Japan},
	number = {12},
	pages = {123705},
	title = {Universal Dynamics of Magnetic Monopoles in Two-Dimensional Kagom{\'e} Ice},
	url = {https://doi.org/10.7566/JPSJ.90.123705},
	volume = {90},
	year = {2021}}

@article{PhysRevB.101.180405,
	author = {Inagaki, K. and Suzuki, S. and Ishikawa, A. and Tsugawa, T. and Aya, F. and Yamada, T. and Tokiwa, K. and Takeuchi, T. and Tamura, R.},
	doi = {10.1103/PhysRevB.101.180405},
	issue = {18},
	journal = {Phys. Rev. B},
	month = {May},
	numpages = {5},
	pages = {180405},
	publisher = {American Physical Society},
	title = {Ferromagnetic 2/1 quasicrystal approximants},
	url = {https://link.aps.org/doi/10.1103/PhysRevB.101.180405},
	volume = {101},
	year = {2020}}

@article{PhysRevB.88.214202,
	author = {Ko\ifmmode \check{z}\else \v{z}\fi{}elj, P. and Jazbec, S. and Vrtnik, S. and Jelen, A. and Dolin\ifmmode \check{s}\else \v{s}\fi{}ek, J. and Jagodi\ifmmode \check{c}\else \v{c}\fi{}, M. and Jagli\ifmmode \check{c}\else \v{c}\fi{}i\ifmmode \acute{c}\else \'{c}\fi{}, Z. and Boulet, P. and de Weerd, M. C. and Ledieu, J. and Dubois, J. M. and Fourn\'ee, V.},
	doi = {10.1103/PhysRevB.88.214202},
	issue = {21},
	journal = {Phys. Rev. B},
	month = {Dec},
	numpages = {11},
	pages = {214202},
	publisher = {American Physical Society},
	title = {Geometrically frustrated magnetism of spins on icosahedral clusters: The Gd${}_{3}$Au${}_{13}$Sn${}_{4}$ quasicrystalline approximant},
	url = {https://link.aps.org/doi/10.1103/PhysRevB.88.214202},
	volume = {88},
	year = {2013}}

@article{Hiroto_2014,
	author = {Hiroto, T and Tokiwa, K and Tamura, R},
	doi = {10.1088/0953-8984/26/21/216004},
	journal = {Journal of Physics: Condensed Matter},
	month = {may},
	number = {21},
	pages = {216004},
	publisher = {IOP Publishing},
	title = {Sign of canted ferromagnetism in the quasicrystal approximants Au-SM-R (SM = Si, Ge and Sn / R = Tb, Dy and Ho)},
	url = {https://doi.org/10.1088/0953-8984/26/21/216004},
	volume = {26},
	year = {2014}}

@article{Hiroto_2013,
	author = {Hiroto, T and Gebresenbut, G H and Pay G{\'o}mez, C and Muro, Y and Isobe, M and Ueda, Y and Tokiwa, K and Tamura, R},
	doi = {10.1088/0953-8984/25/42/426004},
	journal = {Journal of Physics: Condensed Matter},
	month = {sep},
	number = {42},
	pages = {426004},
	publisher = {IOP Publishing},
	title = {Ferromagnetism and re-entrant spin-glass transition in quasicrystal approximants Au--SM--Gd (SM = Si, Ge)},
	url = {https://doi.org/10.1088/0953-8984/25/42/426004},
	volume = {25},
	year = {2013}}

@article{PhysRevB.100.180409,
	author = {Yoshida, S. and Suzuki, S. and Yamada, T. and Fujii, T. and Ishikawa, A. and Tamura, R.},
	doi = {10.1103/PhysRevB.100.180409},
	issue = {18},
	journal = {Phys. Rev. B},
	month = {Nov},
	numpages = {5},
	pages = {180409},
	publisher = {American Physical Society},
	title = {Antiferromagnetic order survives in the higher-order quasicrystal approximant},
	url = {https://link.aps.org/doi/10.1103/PhysRevB.100.180409},
	volume = {100},
	year = {2019}}

@article{PhysRevB.98.220403,
	author = {Ishikawa, A. and Fujii, T. and Takeuchi, T. and Yamada, T. and Matsushita, Y. and Tamura, R.},
	doi = {10.1103/PhysRevB.98.220403},
	issue = {22},
	journal = {Phys. Rev. B},
	month = {Dec},
	numpages = {6},
	pages = {220403},
	publisher = {American Physical Society},
	title = {Antiferromagnetic order is possible in ternary quasicrystal approximants},
	url = {https://link.aps.org/doi/10.1103/PhysRevB.98.220403},
	volume = {98},
	year = {2018}}

@article{PhysRevB.93.024416,
	author = {Ishikawa, A. and Hiroto, T. and Tokiwa, K. and Fujii, T. and Tamura, R.},
	doi = {10.1103/PhysRevB.93.024416},
	issue = {2},
	journal = {Phys. Rev. B},
	month = {Jan},
	numpages = {6},
	pages = {024416},
	publisher = {American Physical Society},
	title = {Composition-driven spin glass to ferromagnetic transition in the quasicrystal approximant Au-Al-Gd},
	url = {https://link.aps.org/doi/10.1103/PhysRevB.93.024416},
	volume = {93},
	year = {2016}}

@article{Ibuka_2011,
	author = {Ibuka, Soshi and Iida, Kazuki and Sato, Taku J},
	doi = {10.1088/0953-8984/23/5/056001},
	journal = {Journal of Physics: Condensed Matter},
	month = {jan},
	number = {5},
	pages = {056001},
	title = {Magnetic properties of the Ag--In--rare-earth 1/1 approximants},
	url = {https://doi.org/10.1088/0953-8984/23/5/056001},
	volume = {23},
	year = {2011}}

@article{doi:10.7566/JPSJ.86.093702,
	author = {Imura ,Keiichiro and Nobe ,Kohei and Deguchi ,Kazuhiko and Matsunami ,Masaharu and Miyazaki ,Hidetoshi and Yasui ,Akira and Ikenaga ,Eiji and Sato ,Noriaki K.},
	doi = {10.7566/JPSJ.86.093702},
	eprint = {https://doi.org/10.7566/JPSJ.86.093702},
	journal = {Journal of the Physical Society of Japan},
	number = {9},
	pages = {093702},
	title = {First Observation of Heavy Fermion Behavior in Ce-Based Icosahedral Approximant},
	url = {https://doi.org/10.7566/JPSJ.86.093702},
	volume = {86},
	year = {2017}}

@article{doi:10.1143/JPSJ.81.024720,
	author = {Mori ,Akinobu and Ota ,Hisashi and Yoshiuchi ,Shingo and Iwakawa ,Ken and Taga ,Yuki and Hirose ,Yusuke and Takeuchi ,Tetsuya and Yamamoto ,Etsuji and Haga ,Yoshinori and Honda ,Fuminori and Settai ,Rikio and {\=O}nuki ,Yoshichika},
	doi = {10.1143/JPSJ.81.024720},
	eprint = {https://doi.org/10.1143/JPSJ.81.024720},
	journal = {Journal of the Physical Society of Japan},
	number = {2},
	pages = {024720},
	title = {Electrical and Magnetic Properties of Quasicrystal Approximants RCd6 (R: Rare Earth)},
	url = {https://doi.org/10.1143/JPSJ.81.024720},
	volume = {81},
	year = {2012}}

@article{PhysRevB.82.220201,
	author = {Tamura, R. and Muro, Y. and Hiroto, T. and Nishimoto, K. and Takabatake, T.},
	doi = {10.1103/PhysRevB.82.220201},
	issue = {22},
	journal = {Phys. Rev. B},
	month = {Dec},
	numpages = {4},
	pages = {220201},
	publisher = {American Physical Society},
	title = {Long-range magnetic order in the quasicrystalline approximant ${\text{Cd}}_{6}\text{Tb}$},
	url = {https://link.aps.org/doi/10.1103/PhysRevB.82.220201},
	volume = {82},
	year = {2010}}

@article{Ramirez:1999aa,
	author = {Ramirez, A. P. and Hayashi, A. and Cava, R. J. and Siddharthan, R. and Shastry, B. S.},
	date = {1999/05/01},
	doi = {10.1038/20619},
	id = {Ramirez1999},
	isbn = {1476-4687},
	journal = {Nature},
	number = {6734},
	pages = {333--335},
	title = {Zero-point entropy in `spin ice'},
	url = {https://doi.org/10.1038/20619},
	volume = {399},
	year = {1999}}

@article{Jaubert:2009aa,
	author = {Jaubert, L. D. C. and Holdsworth, P. C. W.},
	date = {2009/04/01},
	doi = {10.1038/nphys1227},
	id = {Jaubert2009},
	isbn = {1745-2481},
	journal = {Nature Physics},
	number = {4},
	pages = {258--261},
	title = {Signature of magnetic monopole and Dirac string dynamics in spin ice},
	url = {https://doi.org/10.1038/nphys1227},
	volume = {5},
	year = {2009}}

@article{vqvs-rbzn,
	author = {Eto, Rintaro and Mochizuki, Masahito and Watanabe, Shinji},
	doi = {10.1103/vqvs-rbzn},
	issue = {2},
	journal = {Phys. Rev. B},
	month = {Jul},
	numpages = {7},
	pages = {L020405},
	publisher = {American Physical Society},
	title = {Predicted versatile topological nodal magnons in Tb-based icosahedral quasicrystal 1/1 approximants},
	url = {https://link.aps.org/doi/10.1103/vqvs-rbzn},
	volume = {112},
	year = {2025}}

@article{PhysRevB.109.184404,
	author = {Watanabe, Shinji},
	doi = {10.1103/PhysRevB.109.184404},
	issue = {18},
	journal = {Phys. Rev. B},
	month = {May},
	numpages = {7},
	pages = {184404},
	publisher = {American Physical Society},
	title = {Dynamical and static structure factors in hedgehog-antihedgehog order in an icosahedral 1/1 approximant crystal},
	url = {https://link.aps.org/doi/10.1103/PhysRevB.109.184404},
	volume = {109},
	year = {2024}}

@article{Watanabe:2021aa,
	author = {Watanabe, Shinji},
	date = {2021/09/03},
	doi = {10.1038/s41598-021-97024-w},
	id = {Watanabe2021},
	isbn = {2045-2322},
	journal = {Scientific Reports},
	number = {1},
	pages = {17679},
	title = {Magnetism and topology in Tb-based icosahedral quasicrystal},
	url = {https://doi.org/10.1038/s41598-021-97024-w},
	volume = {11},
	year = {2021}}

@article{doi:10.7566/JPSJ.85.053701,
	author = {Sugimoto ,Takanori and Tohyama ,Takami and Hiroto ,Takanobu and Tamura ,Ryuji},
	doi = {10.7566/JPSJ.85.053701},
	eprint = {https://doi.org/10.7566/JPSJ.85.053701},
	journal = {Journal of the Physical Society of Japan},
	number = {5},
	pages = {053701},
	title = {Phenomenological Magnetic Model in Tsai-Type Approximants},
	url = {https://doi.org/10.7566/JPSJ.85.053701},
	volume = {85},
	year = {2016}}

@article{Jeon:2024aa,
	author = {Jeon, Junmo and Lee, SungBin},
	date = {2024/01/06},
	doi = {10.1038/s41535-023-00617-z},
	id = {Jeon2024},
	isbn = {2397-4648},
	journal = {npj Quantum Materials},
	number = {1},
	pages = {5},
	title = {Unveiling multipole physics and frustration of icosahedral magnetic quasicrystals},
	url = {https://doi.org/10.1038/s41535-023-00617-z},
	volume = {9},
	year = {2024}}

@article{PhysRevB.68.024203,
	author = {G\'omez, Cesar Pay and Lidin, Sven},
	doi = {10.1103/PhysRevB.68.024203},
	issue = {2},
	journal = {Phys. Rev. B},
	month = {Jul},
	numpages = {9},
	pages = {024203},
	publisher = {American Physical Society},
	title = {Comparative structural study of the disordered $M{\mathrm{Cd}}_{6}$ quasicrystal approximants},
	url = {https://link.aps.org/doi/10.1103/PhysRevB.68.024203},
	volume = {68},
	year = {2003}}

@article{Takakura:2007aa,
	author = {Takakura, Hiroyuki and G{\'o}mez, Cesar Pay and Yamamoto, Akiji and De Boissieu, Marc and Tsai, An Pang},
	date = {2007/01/01},
	doi = {10.1038/nmat1799},
	id = {Takakura2007},
	isbn = {1476-4660},
	journal = {Nature Materials},
	number = {1},
	pages = {58--63},
	title = {Atomic structure of the binary icosahedral Yb--Cd quasicrystal},
	url = {https://doi.org/10.1038/nmat1799},
	volume = {6},
	year = {2007}}

@article{mizoguchi2018magnetic,
	author = {Mizoguchi, Tomonari and Jaubert, Ludovic D. C. and Moessner, Roderich and Udagawa, Masafumi},
	doi = {10.1103/PhysRevB.98.144446},
	issue = {14},
	journal = {Phys. Rev. B},
	month = {Oct},
	numpages = {22},
	pages = {144446},
	publisher = {American Physical Society},
	title = {Magnetic clustering, half-moons, and shadow pinch points as signals of a proximate Coulomb phase in frustrated Heisenberg magnets},
	url = {https://link.aps.org/doi/10.1103/PhysRevB.98.144446},
	volume = {98},
	year = {2018}}

@article{ANDERSON1978291,
	author = {P.W. Anderson},
	doi = {https://doi.org/10.1016/0022-5088(78)90040-1},
	issn = {0022-5088},
	journal = {Journal of the Less Common Metals},
	pages = {291-294},
	title = {The concept of frustration in spin glasses},
	url = {https://www.sciencedirect.com/science/article/pii/0022508878900401},
	volume = {62},
	year = {1978}}

@book{UdagawaJaubert2021,
	doi = {10.1007/978-3-030-70860-3},
	editor = {Udagawa, Masafumi and Jaubert, Ludovic},
	isbn = {978-3-030-70858-0 978-3-030-70860-3},
	publisher = {Springer International Publishing},
	series = {{Springer Series in Solid-State Sciences}},
	title = {{Spin Ice}},
	url = {https://link.springer.com/10.1007/978-3-030-70860-3},
	volume = {197},
	year = {2021}}

@article{PhysRevB.68.064411,
	author = {Moessner, R. and Sondhi, S. L.},
	doi = {10.1103/PhysRevB.68.064411},
	issue = {6},
	journal = {Phys. Rev. B},
	month = {Aug},
	numpages = {12},
	pages = {064411},
	publisher = {American Physical Society},
	title = {Theory of the [111] magnetization plateau in spin ice},
	url = {https://link.aps.org/doi/10.1103/PhysRevB.68.064411},
	volume = {68},
	year = {2003}}

@article{doi:10.1143/JPSJ.71.2365,
	author = {Udagawa ,Masafumi and Ogata ,Masao and Hiroi ,Zenji},
	doi = {10.1143/JPSJ.71.2365},
	eprint = {https://doi.org/10.1143/JPSJ.71.2365},
	journal = {Journal of the Physical Society of Japan},
	number = {10},
	pages = {2365-2368},
	title = {Exact Result of Ground-State Entropy for Ising Pyrochlore Magnets under a Magnetic Field along [111] Axis},
	url = {https://doi.org/10.1143/JPSJ.71.2365},
	volume = {71},
	year = {2002}}

@article{PhysRevLett.122.117201,
	author = {Udagawa, Masafumi and Moessner, Roderich},
	doi = {10.1103/PhysRevLett.122.117201},
	issue = {11},
	journal = {Phys. Rev. Lett.},
	month = {Mar},
	numpages = {6},
	pages = {117201},
	publisher = {American Physical Society},
	title = {Spectrum of Itinerant Fractional Excitations in Quantum Spin Ice},
	url = {https://link.aps.org/doi/10.1103/PhysRevLett.122.117201},
	volume = {122},
	year = {2019}}

@article{PhysRevLett.124.097204,
	author = {Morampudi, Siddhardh C. and Wilczek, Frank and Laumann, Chris R.},
	doi = {10.1103/PhysRevLett.124.097204},
	issue = {9},
	journal = {Phys. Rev. Lett.},
	month = {Mar},
	numpages = {6},
	pages = {097204},
	publisher = {American Physical Society},
	title = {Spectroscopy of Spinons in Coulomb Quantum Spin Liquids},
	url = {https://link.aps.org/doi/10.1103/PhysRevLett.124.097204},
	volume = {124},
	year = {2020}}

@article{ANDERSON1973153,
	author = {P.W. Anderson},
	doi = {https://doi.org/10.1016/0025-5408(73)90167-0},
	issn = {0025-5408},
	journal = {Materials Research Bulletin},
	number = {2},
	pages = {153 - 160},
	title = {Resonating valence bonds: A new kind of insulator?},
	url = {http://www.sciencedirect.com/science/article/pii/0025540873901670},
	volume = {8},
	year = {1973}}

@article{tokiwa2016tokiwa,
	author = {Tokiwa, Y. and Yamashita, T. and Udagawa, M. and Kittaka, S. and Sakakibara, T. and Terazawa, D. and Shimoyama, Y. and Terashima, T. and Yasui, Y. and Shibauchi, T. and Matsuda, Y.},
	journal = {Nat. Commun.},
	pages = {10807},
	title = {Thermal conductivity of quantum magnetic monopoles in the frustrated pyrochlore Yb2Ti2O7},
	volume = {7},
	year = {2016}}

@article{kitaev2006anyons,
	author = {Kitaev, Alexei},
	journal = {Annals of Physics},
	number = {1},
	pages = {2--111},
	publisher = {Elsevier},
	title = {Anyons in an exactly solved model and beyond},
	volume = {321},
	year = {2006}}

@article{PhysRevLett.119.077207,
	author = {Mizoguchi, Tomonari and Jaubert, L. D. C. and Udagawa, Masafumi},
	doi = {10.1103/PhysRevLett.119.077207},
	issue = {7},
	journal = {Phys. Rev. Lett.},
	month = {Aug},
	numpages = {6},
	pages = {077207},
	publisher = {American Physical Society},
	title = {Clustering of Topological Charges in a Kagome Classical Spin Liquid},
	url = {https://link.aps.org/doi/10.1103/PhysRevLett.119.077207},
	volume = {119},
	year = {2017}}

@article{castelnovo2008magnetic,
	author = {Castelnovo, Claudio and Moessner, Roderich and Sondhi, Shivaji L},
	journal = {Nature},
	number = {7174},
	pages = {42},
	publisher = {Nature Publishing Group},
	title = {Magnetic monopoles in spin ice},
	volume = {451},
	year = {2008}}

@article{PhysRevB.94.104416,
	author = {Udagawa, M. and Jaubert, L. D. C. and Castelnovo, C. and Moessner, R.},
	doi = {10.1103/PhysRevB.94.104416},
	issue = {10},
	journal = {Phys. Rev. B},
	month = {Sep},
	numpages = {24},
	pages = {104416},
	publisher = {American Physical Society},
	title = {Out-of-equilibrium dynamics and extended textures of topological defects in spin ice},
	url = {https://link.aps.org/doi/10.1103/PhysRevB.94.104416},
	volume = {94},
	year = {2016}}

@article{rau2016spin,
	author = {Rau, Jeffrey G and Gingras, Michel JP},
	journal = {Nature communications},
	pages = {12234},
	publisher = {Nature Publishing Group},
	title = {Spin slush in an extended spin ice model},
	volume = {7},
	year = {2016}}

\end{document}